\documentclass[aps,prd,groupedaddress]{revtex4-2}
\pdfoutput=1
\usepackage{graphicx}
\usepackage{fixltx2e}
\usepackage{slashed}
\usepackage[dvipsnames,table]{xcolor}
\usepackage{hyperref} 
\usepackage{xspace}
\usepackage[tight]{subfigure}
\usepackage{amsmath}
\usepackage{amssymb}
\usepackage{amsfonts}
\usepackage{mathrsfs}
\usepackage{comment}
\usepackage{afterpage}
\usepackage{verbatim}
\usepackage{booktabs}
\usepackage{array}
\usepackage[title,titletoc]{appendix}
\usepackage{tabularx}
\newcolumntype{C}[1]{>{\centering\arraybackslash}p{#1}}

\def\refeq#1{\mbox{(\ref{#1})}}
\def\reffi#1{\mbox{Figure~\ref{#1}}}
\def\reffis#1{\mbox{Figures~\ref{#1}}}
\def\refta#1{\mbox{Table~\ref{#1}}}

\def\refse#1{\mbox{Section~\ref{#1}}}

\def\citere#1{\mbox{Ref.~\cite{#1}}}
\def\citeres#1{\mbox{Refs.~\cite{#1}}}

\newcommand{\newc}{\newcommand}
\newc{\beq}{\begin{equation}}
\newc{\eeq}{\end{equation}}
\newc{\bit}{\begin{itemize}}
\newc{\eit}{\end{itemize}}
\newc{\ben}{\begin{enumerate}}
\newc{\een}{\end{enumerate}}
\newc{\bce}{\begin{center}}
\newc{\ece}{\end{center}}
\newc{\bfi}{\begin{figure}}
\newc{\efi}{\end{figure}}

\newcommand{\rT}{{\mathrm{T}}}
\newcommand{\rR}{{\mathrm{R}}}
\newcommand{\rL}{{\mathrm{L}}}
\newcommand{\rF}{{\mathrm{F}}}

\newcommand{\ie}{\emph{i.e.}\ }
\newcommand{\eg}{\emph{e.g.}\ }

\newcommand{\GeV}{\ensuremath{\,\text{GeV}}\xspace}
\newcommand{\TeV}{\ensuremath{\,\text{TeV}}\xspace}

\newcommand{\Pj}{\ensuremath{\text{j}}\xspace}
\newcommand{\Pp}{\ensuremath{\text{p}}}
\newcommand{\Pe}{\ensuremath{\text{e}}\xspace}

\newcommand{\Pq}{\ensuremath{q}}

\newcommand{\Pu}{\ensuremath{\text{u}}\xspace}
\newcommand{\Pd}{\ensuremath{\text{d}}\xspace}

\newcommand{\Pg}{\ensuremath{\text{g}}}

\newcommand{\PW}{\ensuremath{\text{W}}\xspace}
\newcommand{\PZ}{\ensuremath{\text{Z}}\xspace}

\newcommand{\MWOS}{\ensuremath{M_\PW^\text{OS}}\xspace}
\newcommand{\MW}{\ensuremath{M_\PW}\xspace}
\newcommand{\MZOS}{\ensuremath{M_\PZ^\text{OS}}\xspace}
\newcommand{\MZ}{\ensuremath{M_\PZ}\xspace}

\newcommand{\GZOS}{\ensuremath{\Gamma_\PZ^\text{OS}}\xspace}

\newcommand{\GWOS}{\ensuremath{\Gamma_\PW^\text{OS}}\xspace}

\newcommand{\GF}{\ensuremath{G_\mu}}
\newcommand{\alphas}{\ensuremath{\alpha_\text{s}}\xspace}

\newcommand{\MVOS}{\ensuremath{M_{V}^\text{OS}}\xspace}%
\newcommand{\GVOS}{\ensuremath{\Gamma_{V}^\text{OS}}\xspace}%

\newcommand{\recola}{{\sc Recola}\xspace}

\newcommand{\mocanlo}{{\sc MoCaNLO}\xspace}
\newcommand{\bbmc}{{\sc BBMC}\xspace}
\newcommand{\collier}{{\sc Collier}\xspace}

\newcolumntype{.}{D{.}{.}{-1}}
\newcolumntype{d}[1]{D{.}{.}{#1}}
\colorlet{tableoverheadcolor}{gray!37.5}
\colorlet{tableheadcolor}{gray!25}
\colorlet{tablerowcolor}{gray!12.5}

\def\draftdate{\relax}
\def\mda{\relax}
\def\mua{\relax}
\def\mla{\relax}
\def\draft{
\def\thtystars{******************************}
\def\sixtystars{\thtystars\thtystars}
\typeout{}
\typeout{\sixtystars**}
\typeout{* Draft mode!
         For final version remove \protect\draft\space in source file *}
\typeout{\sixtystars**}
\typeout{}
\def\draftdate{\today}
\def\mua{\marginpar[\boldmath\hfil$\uparrow$]%
                   {\boldmath$\uparrow$\hfil}\color{black}%
                    \typeout{marginpar: $\uparrow$}\ignorespaces}
\def\mda{\color{red}\marginpar[\boldmath\hfil$\downarrow$]%
                   {\boldmath$\downarrow$\hfil}%
                    \typeout{marginpar: $\downarrow$}\ignorespaces}
\def\mla{\marginpar[\boldmath\hfil$\rightarrow$]%
                   {\boldmath$\leftarrow $\hfil}%
                    \typeout{marginpar: $\leftrightarrow$}\ignorespaces}
\def\Mua{\marginpar[\boldmath\hfil$\Uparrow$]%
                   {\boldmath$\Uparrow$\hfil}\color{black}%
                    \typeout{marginpar: $\uparrow$}\ignorespaces}
\def\Mda{\color{red}\marginpar[\boldmath\hfil$\Downarrow$]%
                   {\boldmath$\Downarrow$\hfil}%
                    \typeout{marginpar: $\downarrow$}\ignorespaces}
\def\Mla{\marginpar[\boldmath\hfil\textcolor{red}{$\Rightarrow$}]%
                   {\boldmath\textcolor{red}{$\Leftarrow $}\hfil}%
                    \typeout{marginpar: $\leftrightarrow$}\ignorespaces}
\overfullrule 5pt
\oddsidemargin 15mm
\marginparwidth 29mm
}

\newcommand{\mc}{\mathcal}
\newcommand{\as}{\alpha_{\textrm{s}}}
\newcommand{\pt}[1]{p_{\rT,{#1}}}

\newcommand{\nnb}{\nonumber}

\newcommand{\dpathreetwo}{$\rm DPA^{(3,2)}$\xspace}

\newcommand{\CM}{{\scalebox{.6}{\rm CM}}}

\begin{document}

          
   \title{NLO QCD corrections to polarised di-boson production in semi-leptonic final states} 
   \author{Ansgar Denner }\email{ansgar.denner@physik.uni-wuerzburg.de}   
   \author{Christoph Haitz}\email{christoph.haitz@physik.uni-wuerzburg.de}   
   \affiliation{Universit\"at W\"urzburg, Institut f\"ur Theoretische Physik und Astrophysik, 97074 W\"urzburg, Germany}  
   \author{Giovanni Pelliccioli}
   \affiliation{Max-Planck-Institut f\"ur Physik, F\"ohringer Ring 6, 80805 M\"unchen, Germany}\email{gpellicc@mpp.mpg.de}

  \begin{abstract}
    Understanding the polarisation structure and providing precise predictions for
    multi-boson processes at the LHC is becoming urgent in the light
    of the upcoming run-3 and high-luminosity data.
    The CMS and ATLAS collaborations have already started using polarised predictions
    to perform template fits of the data, getting access to the polarisation of $\PW$
    and $\PZ$~bosons. So far, only
    fully-leptonic decay channels have been considered in this perspective. The natural
    step forward is the investigation of hadronic decays of electroweak bosons.
    In this work, we compute NLO QCD corrections to the production and decay of WZ
    pairs at the LHC in final states with two charged leptons and jets. The calculation
    relies on the double-pole approximation and the separation of polarised states at the
    level of Standard Model amplitudes. 
    The presented NLO-accurate results are necessary building blocks for
    a broad understanding and precise modelling of polarised di-boson production in
    semi-leptonic decay channels.
  \end{abstract}

\keywords{polarisation, electroweak bosons, NLO QCD, semi-leptonic, LHC}

\preprint{MPP-2022-135}
          
          

\maketitle
\tableofcontents
\section{Introduction}\label{sec:intro}

The pioneering measurements  with the run-2 LHC dataset in di-boson inclusive production
\cite{Aaboud:2019gxl,CMS:2021icx,ATLAS:2022oge} and scattering (VBS) \cite{Sirunyan:2020gvn}
have paved the way towards a refined experimental investigation of the polarisation
states of electroweak (EW) bosons produced in multi-boson processes.
The most striking difference of the methods adopted in \citeres{Aaboud:2019gxl,CMS:2021icx,ATLAS:2022oge,Sirunyan:2020gvn}
with respect to previous polarisation analyses
\cite{Chatrchyan:2011ig,ATLAS:2012au,ATLAS:2016fbc,Khachatryan:2016fky,Khachatryan:2015paa,Aad:2016izn,CMS:2020ezf}
is the usage of polarised-signal templates directly generated with Monte Carlo generators,
which is expected to give a more complete picture of the polarisation structure than the simple
evaluation of angular coefficients, giving access to interference effects and spin correlations.

The extraction of angular coefficients from unpolarised decay-angle distributions
was first proposed in \citere{Bern:2011ie,Stirling:2012zt} and has been applied in
several theoretical studies on $V+\Pj$ 
\cite{Bern:2011ie,Stirling:2012zt,Gauld:2017tww,Frederix:2020nyw,Pellen:2022fom}
and inclusive di-boson production
\cite{Rahaman:2018ujg,Rahaman:2019lab,Baglio:2018rcu,Baglio:2019nmc,Rahaman:2021fcz}.
The direct Monte Carlo simulation of processes with intermediate EW polarised bosons
was proposed for VBS \cite{Ballestrero:2017bxn,Ballestrero:2019qoy,Ballestrero:2020qgv,BuarqueFranzosi:2019boy}
and it is currently available in public codes at leading-order (LO) accuracy matched to
parton-shower (PS) in the Standard Model (SM) as well
as in the presence of beyond-SM effects. The extension of this approach to higher perturbative
orders has been carried out focusing on di-boson inclusive production, reaching
next-to-leading-order (NLO) in the QCD coupling \cite{Denner:2020bcz,Denner:2020eck}
and later on NLO in the EW coupling \cite{Denner:2021csi,Le:2022lrp,Le:2022ppa}
and next-to-next-to-leading-order (NNLO) in QCD \cite{Poncelet:2021jmj}.
Analogous methods have been applied to Higgs decays \cite{Maina:2020rgd,Maina:2021xpe}
and to $V+\Pj$~production \cite{Pellen:2021vpi}.
A number of works have tackled the polarisation structure of VBS with machine-learning
techniques with promising results \cite{Searcy:2015apa,Lee:2018xtt,Lee:2019nhm,Grossi:2020orx}.

So far, both experimental measurements
\cite{Chatrchyan:2011ig,ATLAS:2012au,Khachatryan:2015paa,ATLAS:2016fbc,Khachatryan:2016fky,Aad:2016izn,
  Aaboud:2019gxl,CMS:2020ezf,CMS:2021icx,ATLAS:2022oge,Sirunyan:2020gvn}
and phenomenological studies \cite{ Han:2009em,
  Bern:2011ie,Stirling:2012zt,Belyaev:2013nla, Brehmer:2014pka,
  Searcy:2015apa,Lee:2018xtt,Lee:2019nhm, Brass:2018hfw,
  Gauld:2017tww,Frederix:2020nyw,Pellen:2022fom,
  Ballestrero:2017bxn,Ballestrero:2019qoy,Ballestrero:2020qgv,
  BuarqueFranzosi:2019boy,
  Denner:2020bcz,Denner:2020eck,Denner:2021csi,
  Poncelet:2021jmj,Pellen:2021vpi,
  Baglio:2018rcu,Baglio:2019nmc,Le:2022lrp,Le:2022ppa, Cao:2020npb,
  Maina:2020rgd,Maina:2021xpe,
  Rahaman:2018ujg,Rahaman:2019lab,Rahaman:2021fcz} have focused on
leptonic decays of weak bosons, cleaner than the hadronic ones in a
hadron-collider environment, but with a smaller branching ratio and
affected by reconstruction effects in the presence of neutrinos in the
final state.  Indeed, the increasing interest in measuring gauge-boson
polarisations at the LHC and the lack of statistics in fully-leptonic
decay channels has triggered a number of phenomenological studies
\cite{De:2020iwq,Kim:2021gtv,Dey:2021sug,Ricci:2022htc} of processes
with hadronically decaying bosons.  The hadronic decay of $\PW$ and
$\PZ$~bosons has the great advantage of larger branching ratio w.r.t.\ 
the leptonic one, but the disadvantage of much larger backgrounds to
deal with in the LHC environment that is dominated by QCD-induced
processes.  The focus of these polarisation studies has been put
especially on the discrimination between longitudinal and transverse
bosons in boosted kinematic configurations
\cite{De:2020iwq,Kim:2021gtv,Dey:2021sug}.  The driving idea is that
the reconstruction of the boosted-fat-jet substructure, making use of
either traditional substructure observables like $N$-subjettiness
\cite{De:2020iwq} and soft-drop \cite{Dey:2021sug}, or
machine-learning \cite{Kim:2021gtv} techniques, is expected to
maximise the information about the polarisation state of the decayed
boson, since the jet constituents can be associated to some degree of
precision to the decay quarks. Recently it has been proposed to use
energy correlators to improve the polarisation discrimination
\cite{Ricci:2022htc}.  The existing studies of polarised bosons in the
leptonic decay channel suggest that there are LHC observables that
enable to discriminate polarisation states without the need to
reconstruct individual decay products \cite{ATLAS:2022oge}.

So far, no polarisation measurement has been carried out yet with
hadronic decays of gauge bosons, although a number of sensitivity
studies have been performed for the high-energy
\cite{Cavaliere:2018zcf} and high-luminosity (HL)
\cite{Roloff:2021kdu} runs of the LHC, mostly targeting VBS processes.

In the specific case of inclusive di-boson production, the semi-leptonic decay channel has been investigated
with 13-TeV LHC data with the aim of searching for new resonances
\cite{CMS:2018sdh,ATLAS:2018sbw,CMS:2019kaf,ATLAS:2020fry,CMS:2021xor}. Measuring polarisations in
this channel could further constrain new-physics effects \cite{Liu:2018pkg}
and in particular the spin of possible underlying resonances decaying in two gauge bosons.
In spite of the high-precision SM predictions available for unpolarised 
\cite{Kallweit:2017khh,Biedermann:2016guo,Biedermann:2016lvg,Biedermann:2017oae,
Cascioli:2014yka,Grazzini:2015hta,Grazzini:2016ctr,Grazzini:2016swo,Heinrich:2017bvg,Grazzini:2017ckn,Kallweit:2018nyv,
Kallweit:2019zez,Caola:2015rqy,Caola:2015psa,Grazzini:2018owa,
Alioli:2016xab,Re:2018vac,Brauer:2020kfv,Chiesa:2020ttl,Alioli:2021egp,Alioli:2021wpn,Lindert:2022qdd}
and polarised \cite{Denner:2020bcz,Denner:2020eck,Poncelet:2021jmj,Denner:2021csi,Le:2022lrp,Le:2022ppa} di-boson production at
the LHC, the fully-leptonic decay channel has always been considered and no tailored study beyond LO (+PS) exists for the
semi-leptonic channel in the presence of polarised intermediate bosons.
The NLO QCD corrections to the hadronic decay of polarised weak bosons have been known for many years 
\cite{Groote:2012xr,Groote:2013xt} and can be generated easily with the help of any one-loop
amplitude provider, but they have not been combined yet with realistic LHC production processes.

In this work we perform a consistent combination of NLO QCD corrections to di-boson production and to the
hadronic decay of one of the two bosons in the double-pole
approximation%
\footnote{For the pole approximation for one resonance see also \protect\citeres{Stuart:1991cc,Stuart:1991xk}.}
\cite{Aeppli:1993cb,Aeppli:1993rs,Denner:2000bj,Denner:2005fg,Denner:2019vbn}
and in the presence of polarised and unpolarised intermediate bosons, preserving partial off-shell effects and
the complete spin correlations between production and decay. This target is pursued following the
strategy proposed at LO in \citere{Ballestrero:2017bxn} and later extended to NLO in \citeres{Denner:2020bcz,Denner:2021csi,Le:2022ppa}.
We consider a boosted regime for the weak bosons, where the doubly-longitudinal signal
is expected to be sizeable \cite{Denner:2020eck}, but we do not make use of any jet-substructure-reconstruction technique.

This paper is organised as follows. In \refse{sec:outline} we describe the technical details of the
calculation, the SM input parameters and the kinematic setups that are considered. In \refse{sec:res} we
present and discuss the integrated and differential predictions for doubly-polarised signals at the LHC with
13.6-TeV centre-of-mass (CM) energy.
Our conclusions and outlook are given in \refse{sec:con}.

\section{Details of the calculation}\label{sec:outline}
We consider di-boson ($\PZ \PW^{+}$) inclusive production at the LHC:
\beq\label{eq:processdef}
  \Pp\Pp\rightarrow \PZ\,(\rightarrow\ell^+\ell^-)\,\PW^{+} \,(\rightarrow{\rm jj})
  + X\,,\qquad
  \Pp\Pp\rightarrow \PZ\,(\rightarrow\ell^+\ell^-)\,\PW^{+} \,(\rightarrow{\rm J}) + X\,,
\eeq
where the two semi-leptonic decay channels differ by the number of jets from the
decay of the W~boson. In the first case (resolved topology)
the $\PW^{+}$ boson decays into two light jets, while in the second case (unresolved
topology) it decays into a single fat jet. We calculate the NLO QCD corrections to the tree-level
EW process, \ie of perturbative order $\mc{O}(\as\alpha^4)$,
working in the double-pole approximation (DPA)
\cite{Aeppli:1993cb,Aeppli:1993rs,Denner:2000bj,Denner:2005fg,Denner:2019vbn}.
The non-resonant contributions and off-shell effects beyond the DPA, as
well as the other perturbative orders contributing to the same final state
(which cannot embed two weak-boson propagators) are not considered here.
In other words, we do not include the irreducible non-resonant and
QCD multi-jet backgrounds to the di-boson signal. We also restrict the calculation
to $\PZ\PW^+$~production, neglecting the $\PZ\PW^-$ and $\PZ\PZ$ resonant processes
which also contribute to the same final state with two leptons and a hadronic system.
  A rough estimate of the missing resonant contributions
  ($\PZ\PW^-$ and $\PZ\PZ$), of the non-resonant EW effects,
  and of the QCD background is provided at the end of \refse{subsec:int_res}.
  Owing to the different resonance and spin structures of the various
  sub-processes in the SM,
  it is preferable to investigate doubly-polarised signals with LHC data
  focusing on one resonant structure at a time, \ie considering other
  (unpolarised) resonant processes as contributions to be
  subtracted on equal footing with the QCD irreducible backgrounds
  and off-shell EW effects.
  This approach could be beneficial also in the lights of possible
  new-physics effects that may distort the SM dynamics differently
  for various di-boson processes.
  Therefore, the study of $\PW^+\PZ$ production presented here, 
which is
not meant to be fully realistic, is expected to give useful insights for future
analyses in semi-leptonic final states with run-3 and HL-LHC data.
  
\subsection{Double-pole approximation and polarised-signal definition}\label{sec:outline_subsec:DPA}
\begin{figure}[tb]
  \centering
  \subfigure[\label{subfig:tree}]{\includegraphics[scale=0.5]    {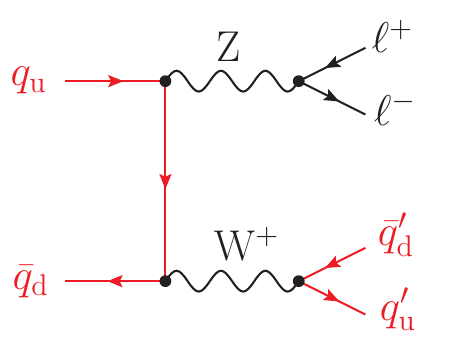}                }
  \subfigure[\label{subfig:loopI}]{\includegraphics[scale=0.5]   {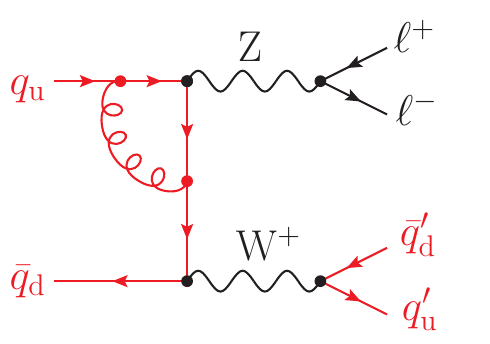}           }
  \subfigure[\label{subfig:loopF}]{\includegraphics[scale=0.5]   {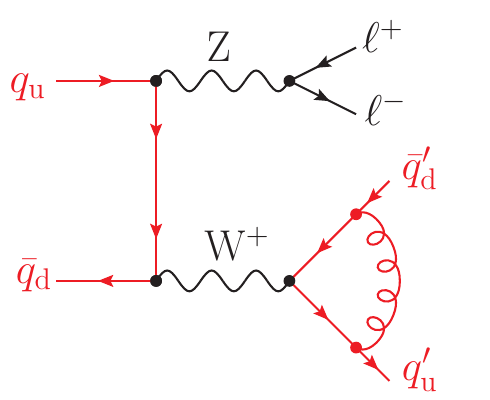}         }
  \subfigure[\label{subfig:isr}]{\includegraphics[scale=0.5]     {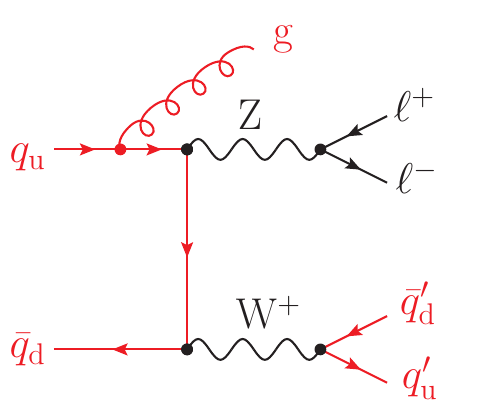}         }
  \subfigure[\label{subfig:isrcross}]{\includegraphics[scale=0.5]{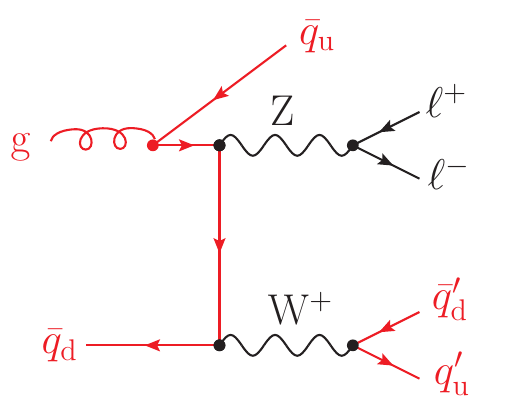} }
  \subfigure[\label{subfig:fsr}]{\includegraphics[scale=0.5]     {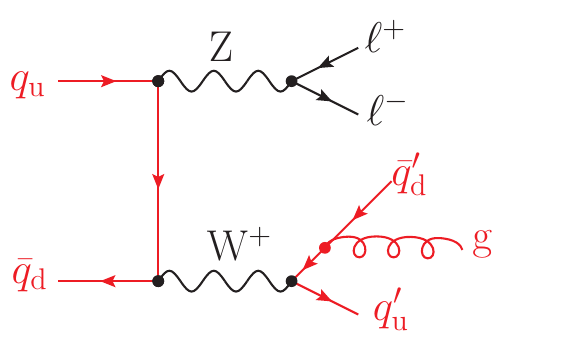}    }
  \caption{Sample tree-level (a), one-loop (b-c), and real-radiation (d-f) diagrams
    contributing to di-boson production at the LHC in the semi-leptonic decay channel
    at NLO QCD. Particles carrying colour charge are highlighted in red.}\label{fig:diags}
\end{figure}

Sample diagrams for $\PZ\PW^+$~production that contribute in the DPA at LO and at NLO QCD are shown in \reffi{fig:diags}.
In the DPA
\cite{Aeppli:1993cb,Aeppli:1993rs,Denner:2000bj,Denner:2005fg,Denner:2019vbn}
all diagrams without two $s$-channel boson propagators (one $\PW^+$
and one $\PZ$~boson) are neglected, giving an amplitude (which is not
gauge invariant) that is fully factorised in a production $\times$
decay form.  In order to recover the EW gauge invariance, the
numerator of the doubly-resonant amplitude is evaluated with a
modified kinematics obtained via an on-mass-shell projection of the
resonant EW bosons, while the denominator (\ie the two Breit--Wigner
distributions from the weak-boson propagators) is evaluated with the
original kinematics (for off-shell bosons).  The widths of the $\PW$
and $\PZ$~bosons are set to zero in the amplitude numerator, while
they are kept finite in the denominators of the weak-boson propagators
targeted by the DPA. A detailed explanation of the DPA and the
gauge invariance in this formalism can be found in
\citeres{Denner:2000bj,Billoni:2013aba,Denner:2019vbn,Denner:2020bcz}.

This technique preserves all spin correlations (the full spin matrix
is accounted for) and partial off-shell effects (thanks to the
Breit--Wigner modulation and the use of the off-shell phase space).
An alternative method that is often used for polarised-signal
simulation
\cite{BuarqueFranzosi:2019boy,Poncelet:2021jmj,Pellen:2021vpi} is the
narrow-width approximation \cite{Uhlemann:2008pm,Artoisenet:2012st}.
While in the narrow-width approximation the off-shell effects are neglected,
the DPA takes into account off-shell effects partially.  Specifically,
the phase space and the Breit--Wigner denominators of the resonances
are treated off shell, and only the matrix-element numerators are
projected on the resonance mass shell.

The DPA approach makes it natural to separate polarisation states of intermediate
weak bosons at amplitude level \cite{Ballestrero:2017bxn,Denner:2020bcz}.
A priori, the polarisation of particles in scattering processes can only be defined
for stable external particles, while for intermediate (virtual) particles it is required
to perform the sum over all polarisations (physical and unphysical).
The factorised structure of double-pole-approximated, doubly-resonant amplitudes enables
the splitting of the numerator of each gauge-boson propagator into the sum over physical
polarisation states: longitudinal (L), left handed ($-$) and right
handed (+). The contributions of unphysical polarisation states are
exactly cancelled by those of would-be Goldstone bosons on the mass
shell. Therefore,
replacing the polarisation sum in the propagator numerator with the
contribution of a specific polarisation state $\lambda$ gives a gauge-invariant
$\lambda$-polarised amplitude, where $\lambda=0,\pm$.
A convenient choice \cite{Ballestrero:2019qoy} is to consider the transverse-polarisation
state (T) which is defined as the coherent sum of the left- and right-handed
states, including also the left--right interference term. 

It is essential to recall that the polarisation vectors appearing in the propagator numerator
depend on the Lorentz frame where the kinematics is evaluated. This implies that the
definition of the polarised signal is reference-frame dependent.
The preferred choice for di-boson inclusive production is the di-boson--system
CM frame \cite{Aaboud:2019gxl,Baglio:2019nmc,Denner:2020eck}, as it
allows for a natural interpretation in terms of the corresponding tree-level
$2\rightarrow 2$ process ($q_{\Pu}\bar{q}_{\Pd}\rightarrow \PZ\PW^+$). This is the
choice adopted in this work.

As previously stated, the novel aspect of the presented calculation with respect to
\citere{Denner:2020eck} is the semi-leptonic decay channel [see Eq.~\eqref{eq:processdef}].
In the fully-leptonic decay channel, the NLO QCD corrections only enter as initial-state
virtual and real radiation. Here we also need to consider the NLO QCD corrections to the
$\PW^+$-boson decay. As can be appreciated in \reffi{fig:diags}, the factorisable virtual
corrections to the decay [\reffi{subfig:loopF}] and to the production sub-process
[\reffi{subfig:loopI}] are accounted for. The real corrections include initial-state-radiation
[\reffis{subfig:isr}--\ref{subfig:isrcross}] and final-state-radiation [\reffi{subfig:fsr}]
contributions. Both virtual and real non-factorisable NLO QCD corrections to doubly-resonant
amplitudes (gluon exchange between production and decay) vanish owing to colour algebra.

The treatment of factorisable real QCD corrections in the DPA is carried out using the
general approach introduced in \citeres{Beenakker:1997ir,Denner:1997ia,Beenakker:1998gr,Denner:2021csi} for NLO EW corrections,
upon the replacement of the EW
coupling $\alpha$ with a running strong coupling $\alphas$.
In order to end up with a final result that is free of infrared (IR) divergences,
unresolved real radiation from the production and decay parts of the
process need to be managed separately in the DPA and combined consistently
within the employed subtraction scheme, in order to have
\begin{itemize}
\item the correct matching between subtraction
counterterms and the corresponding integrated ones, and
\item the cancellation of IR poles between
the integrated counterterms and the virtual matrix element.
\end{itemize}
In particular, gluon radiation from the $\PW$-boson decay products
and from the $\PW\PZ$-pair-production process are treated separately within the DPA.
In our calculation, we employ the dipole subtraction scheme \cite{Catani:1996vz,Dittmaier:1999mb,Catani:2002hc}.
For additional technical details, we refer to \citere{Denner:2021csi}.

\subsection{Monte Carlo tools and input parameters}\label{sec:kinsetups}
The presented SM calculation has been performed with two independent
in-house multi-channel Monte Carlo codes, \mocanlo and \bbmc.
\mocanlo has been recently employed for NLO EW and QCD corrections
to processes with polarised EW bosons in the fully-leptonic decay channel
\cite{Denner:2020bcz,Denner:2020eck,Denner:2021csi}. \bbmc has been used
for NLO corrections to off-shell di-boson production
\cite{Biedermann:2016guo,Biedermann:2016yvs,Biedermann:2016lvg,Biedermann:2017oae}
and has been modified to enable the treatment of resonances in the DPA.
The UV-renormalised tree-level and one-loop SM amplitudes are provided to both codes by
\recola \cite{Actis:2012qn,Actis:2016mpe}. The tensor reduction and integration of
loop integrals is performed with \collier \cite{Denner:2016kdg}.

The calculation is carried out in the SM at NLO QCD accuracy. The
five-flavour scheme and no quark-family mixing (unit CKM matrix) are understood.
All light quarks and leptons are considered massless. 
The pole masses ($M_V$) and widths ($\Gamma_V$) of the EW bosons are calculated from the
on-shell values ($\MVOS,\,\GVOS$) \cite{Tanabashi:2018oca},
\begin{alignat}{2}\label{eq:ewmasses}
 \MWOS &= 80.379 \GeV,&\qquad \GWOS &= 2.085\GeV, \nnb\\
 \MZOS &= 91.1876 \GeV,&\qquad \GZOS &= 2.4952\GeV, 
\end{alignat}
according to the relations \cite{Bardin:1988xt}
\beq
 M_V = \frac{\MVOS}{\sqrt{1+(\GVOS/\MVOS)^2}}\,,\qquad  
 \Gamma_V = \frac{\GVOS}{\sqrt{1+(\GVOS/\MVOS)^2}}.
\eeq
The EW coupling $\alpha$ is calculated in the $G_\mu$ scheme \cite{Denner:2000bj},
\ie as a function of the Fermi constant $\GF$ and the weak-boson pole masses:
\beq
\alpha = \frac{\sqrt{2}}{\pi}\,G_\mu\MW^2\left(1-\frac{\MW^2}{\MZ^2}\right)\,,\qquad\,
\GF = 1.16638\cdot 10^{-5} \GeV^{-2}.
\eeq
Since we only consider NLO QCD corrections,
the masses of the top quark and of the Higgs boson
do not enter the calculation.
We use \sloppy\textsc{NNPDF31\_nlo\_as\_0118} \cite{Ball:2017nwa,Bertone:2017bme}
parton-distribution functions (PDFs), provided to the Monte Carlo codes via the LHAPDF interface
\cite{Buckley:2014ana}. Also the running of the strong coupling constant $\alphas$
is evaluated with built-in LHAPDF routines.
The dipole formalism \cite{Catani:1996vz,Dittmaier:1999mb,Catani:2002hc}
is used for the subtraction of IR singularities of QCD origin.
The $\overline{\rm MS}$ factorisation scheme is employed for the treatment of
initial-state collinear singularities.
The central factorisation and renormalisation scales are
both set to the same dynamical value $\mu_{\rm F}=\mu_{\rm R}=\mu_0 $
(defined in \refse{subsec:setupscale}), and the 
QCD-scale uncertainties are estimated with independent 7-point
variations of $\mu_{\rm F}$ and $\mu_{\rm R}$, \ie
via the maxima and minima of the corresponding observables for the
scale choices
$$
(\mu_{\rm R}/\mu_0,\mu_{\rm F}/\mu_0) = 
(1/2,1/2), (1/2,1),(1,1/2),(1,1)(1,2),(2,1),(2,2)\,. 
$$

\subsection{Kinematic setups and scale definition}\label{subsec:setupscale}
The event selection and reconstruction are inspired by the recent CMS analysis presented in \citere{CMS:2021xor} (see Table 1 therein).
The clustering of jets is carried out with the anti-$k_{\rm T}$ algorithm \cite{Cacciari:2008gp} recombining only partons with
a rapidity smaller than~5. Two different event topologies are considered: \emph{resolved} and \emph{unresolved}.
In the resolved topology we ask for:
\begin{itemize}
\item at least two jets (clustered with $R_0=0.4$) with
  $\pt{\rm j}>30\GeV$, $|y_{\rm j}|<2.4$ and $\Delta R_{\rm j
    \ell^{\pm}}>0.4$, 
the two jets with a pair invariant mass closest to $\MW$ being selected
as \emph{decay jets}, 
\item  the system of the two decay jets with
  $\pt{\rm jj}>200\GeV$ 
  and $65\GeV < M_{\rm jj}<105\GeV$; 
\item two opposite-sign, same-flavour leptons with
  $\pt{\ell^\pm}>40\GeV$, $|y_{\ell^\pm}|<2.4$, $\pt{\ell\ell}>200\GeV$ 
  and
  $76\GeV<M_{\ell\ell}<106\GeV$.
\end{itemize}
In the unresolved topology we ask for:
\begin{itemize}
\item at least one jet (clustered with $R_0=0.8$) with $|y_{\rm J}|<2.4$, $\Delta R_{\rm J \ell^{\pm}}>0.8$, $\pt{\rm J}>200\GeV$ and $65\GeV < M_{\rm J}<105\GeV$,
\item two opposite-sign, same-flavour leptons with
  $\pt{\ell^\pm}>40\GeV$, $|y_{\ell^\pm}|<2.4$, $\pt{\ell\ell}>200\GeV$ and
  $76\GeV<M_{\ell\ell}<106\GeV$.
\end{itemize}
By requiring large transverse momenta for the vector bosons, we select
a boosted regime.
We do not apply any veto on additional jets,
as the logarithmically-en\-hanced soft-boson radiation in real contributions
\cite{Rubin:2010xp,Baglio:2013toa,Kallweit:2019zez} is suppressed thanks to
the tight transverse-momentum cuts on the two bosons
($\pt{\ell\ell}>200\GeV$ and $\pt{\rm J/jj}>200\GeV$), avoiding huge
QCD $K$-factors.
Note, however, that the application of symmetric transverse-momentum cuts on the two bosons
leads to very large corrections in transverse-momentum distributions close
to the cut, due to the sensitivity to quasi-soft and quasi-collinear
QCD initial-state radiation 
\cite{Klasen:1995xe,Harris:1997hz,Frixione:1997ks,Denner:2011id,Salam:2021tbm}.\\
The central renormalisation and factorisation scales are set to,
\beq\label{eq:scaledef}
\mu_{\rR}=\mu_{\rF}=\frac{M_{\rT, \PZ}+M_{\rT, \rm J}}2\,.
\eeq
The transverse masses are calculated as,
\beq\label{eq:mtdef}
M_{\rT, \PZ} = \sqrt{\pt{\ell\ell}^2+M^2_{\ell\ell}}\,,
\qquad M_{\rT, \rm J} = \sqrt{\pt{\rm J}^2+M^2_{\rm J}}\,,
\eeq
where $\pt{\rm J}$ and $M_{\rm J}$ are, respectively, the transverse momentum and
invariant mass of the hadronic system~$\rm J$ that is identified as the
decay-jet system (resolved topology) or as the hardest-$p_{\rT}$ fat jet (unresolved topology).

As previously stated, the polarisation states of intermediate gauge
bosons are defined in the CM frame, \ie the rest
frame of the system formed by the two leptons and the decay products
of the $\PW$~boson, which are identified for each contribution in the
DPA. This choice enables for qualitative comparisons with
phenomenological studies of $\PZ\PW$~production in the fully-leptonic
decay channel \cite{Denner:2020eck,Le:2022lrp}.

\subsection{Validation}
The independent implementation in \mocanlo and \bbmc of the general methods described in
\citere{Denner:2021csi} has enabled validation checks at several levels of the calculation.
\paragraph{Renormalisation and phase-space integration} Detailed comparisons have been performed both for
individual phase-space points for Born-level, virtual and real-radiation contributions,
giving excellent agreement both for the QCD-scale and matrix-element evaluation.
The UV finiteness of virtual amplitudes in the presence of polarised intermediate
bosons has been checked varying by several orders of magnitude the UV-scale regulator
as input parameter for \recola. Note that both codes make use of
\recola as amplitude provider, which has been successfully
  compared earlier to other one-loop providers for unpolarised
  vector-boson pair production processes 
\cite{Bendavid:2018nar} and to an independent in-house code for WZ
production with leptonic decays \cite{Biedermann:2017oae}.
In addition, we have performed a check of the finite part of virtual QCD corrections
to $q_{\Pu} \bar{q}_{\Pd}\rightarrow \Pe^+ \Pe^- q_{\Pu}' \bar{q}_{\Pd}'$, comparing
the \recola results against a stand-alone version of {\sc MadLoop} \cite{Hirschi:2011pa}.
A point-wise agreement at the $10^{-7}$ level was found for a number of phase-space points
populating both the bulk and far off-shell regions of the fiducial volume.

The correct application of selection cuts (after jet clustering)
in the two setups described in \refse{subsec:setupscale} has been validated comparing results
from \mocanlo and \bbmc for each $n$-body and $(n+1)$-body contribution to the cross-section separately.
This has been done both for the unpolarised and the polarised process, finding agreement within
Monte Carlo uncertainties in all cases.
\paragraph{Subtraction of IR singularities} The correct subtraction of IR singularities of QCD origin
has been tested in depth, with both comparisons between the two codes
and internal checks in each code separately.  The dipole-subtraction
kernels and the kinematics used to evaluate them has been tested for a
number of individual phase-space points, finding good agreement
between the two codes. This is especially relevant for subtraction
dipoles associated to decay sub-process (gluon radiation off the
$\PW$~decay), whose kinematics has to undergo first the \dpathreetwo
on-shell projection and second the final-state--final-state
Catani--Seymour mapping \cite{Denner:2021csi}.  Analogous checks have
been performed on integrated subtraction counterterms, finding also
excellent agreement for each phase-space point considered.  The proper
cancellation of IR poles in the sum of virtual matrix elements and
$I$-operators \cite{Catani:1996vz} has been successfully checked
varying the IR-regularisation scale $\mu_{\rm IR}$ up and down by 4
orders of magnitude about the scale defined in
Eq.~\eqref{eq:scaledef}, finding independence of the result from
$\mu_{\rm IR}$ within the errors of the Monte Carlo integration.  The
functioning of the subtraction scheme in the DPA has been also tested
by means of the technical parameters $\{\alpha_{\rm dip}\}$ that set
the integration boundaries for the radiation phase space in each
dipole \cite{Nagy:1998bb}. Varying such parameters between 1 and
$10^{-2}$, has enabled to check that the sum of subtracted-real and
integrated-counterterm contributions is independent of the choice of
$\{\alpha_{\rm dip}\}$ within integration uncertainties.  The
independence of the NLO corrections from the unphysical parameters
$\mu_{\rm IR}$ and $\{\alpha_{\rm dip}\}$ represents a strong
check given the different treatment of real radiation from the
production and decay sub-process matrix elements and subtraction
counterterms in the DPA. This further confirms that the methods
introduced in \citere{Denner:2021csi} provide NLO predictions that are
well under control from the IR-subtraction point of view.

\paragraph{Treatment of polarisations} The definition of polarisation vectors in the CM reference frame
represents a crucial step in the calculation, especially for what concerns real radiation.
Detailed checks at the phase-space-point level have been done in the
Lorentz frame where polarisations are defined 
in the case of real corrections to the production and
decay sub-processes.
A complete comparison between \mocanlo and \bbmc has been performed for each contribution to the cross-section,
for all doubly-polarised states, giving excellent agreement within integration errors for both
integrated result and differential distributions (bin by bin).
Further tests have been carried out comparing the two codes for a different definition of polarisation,
\ie in the laboratory frame, finding also agreement.

\section{Results}\label{sec:res}
In this section we present numerical results at integrated and differential level for
doubly-polarised and unpolarised $\PZ\PW^+$~production at the LHC in the semi-leptonic
channel. All results shown have been obtained with \mocanlo. We have considered the specific
case of an electron--positron pair ($\ell = \Pe$). The sum over light lepton flavours
($\ell = \Pe, \mu$) can be simply obtained upon multiplying all
cross-sections by a factor of two.

\subsection{Integrated Results}
\label{subsec:int_res}
In \refta{table:sigmainclNLO} we present integrated cross-sections for different polarisation states
in the two setups introduced in \refse{subsec:setupscale}.

\begin{table*}[tb]
\begin{center}
\renewcommand{\arraystretch}{1.3}
\begin{tabular}{C{1.7cm}C{2.3cm}C{1.6cm}C{2.3cm}C{1.6cm}C{1.3cm}C{1.3cm}}%
\hline %
\cellcolor{blue!14} state  & \cellcolor{blue!14}  $\sigma_{\rm LO}$ [fb]  & \cellcolor{blue!14} $f_{\rm LO}[\%]$& \cellcolor{blue!14} $\sigma_{\rm NLO}$ [fb] & \cellcolor{blue!14} f$_{\rm NLO}[\%]$ & \cellcolor{blue!14} {$K_{\rm NLO}$} & \cellcolor{blue!14} {$K_{\rm NLO}^{\rm (no\,g)}$}      \\
\hline
\multicolumn{7}{l}{\cellcolor{green!9} resolved, $\PZ(\Pe^+\Pe^-)\PW^+(\Pj\Pj)$} \\
\hline
unpol. & $  1.8567  ( 2 )^{+ 1.2 \%}_{- 1.4 \%}$ &$100$&        $ 3.036(2)  ^{+ 6.8 \%}_{- 5.3 \%} $ &$ 100 $ & $         1.635 $  & $ 1.033$\\
$\PZ^{\,}_{\rL}\PW^{+}_{\rL}$& $  0.64603  ( 5 ) ^{+ 0.2 \%}_{- 0.6 \%}$  &$ 34.8$&      $ 0.6127  ( 4 ) ^{+ 0.9 \%}_{- 0.7 \%} $  &$ 20.2 $ & $        0.948 $  & $1.031 $\\
$\PZ^{\,}_{\rL}\PW^{+}_{\rT}$& $  0.08687  ( 1 ) ^{+ 0.2 \%}_{- 0.6 \%}$  &$ 4.7$&       $ 0.17012  ( 6 ) ^{+ 8.6 \%}_{- 6.8 \%} $  & $ 5.6 $& $          1.958 $  & $ 0.967 $\\
$\PZ^{\,}_{\rT}\PW^{+}_{\rL}$& $  0.08710  ( 1 ) ^{+ 0.1 \%}_{- 0.6 \%}$  &$ 4.7$&       $ 0.24307  ( 7 ) ^{+ 10.2 \%}_{- 8.2 \%} $  &$ 8.0 $ & $         2.791  $ & $1.017 $ \\
$\PZ^{\,}_{\rT}\PW^{+}_{\rT}$& $  0.97678  ( 7 ) ^{+ 2.0 \%}_{- 2.2 \%}$  &$ 52.6$&      $ 2.0008  ( 7 ) ^{+ 8.9 \%}_{- 7.1 \%} $  &$ 65.8 $ & $        2.048 $  & $1.059 $\\
interf. & $ 0.0595(1) $  &$3.2 $&      $ 0.009(2)$  &$0.4 $ & $-$ & $- $\\
\hline
\multicolumn{7}{l}{\cellcolor{green!9}  unresolved, $\PZ(\Pe^+\Pe^-)\PW^+({\rm J})$} \\
\hline
unpol. &$  1.6879  ( 2 )^{+ 1.9 \%}_{- 2.1 \%}$  &$100$&        $ 3.112(2)  ^{+ 7.6 \%}_{- 6.1 \%} $ &$100 $  & $         1.843 $ & $ 1.193$ \\
$\PZ^{\,}_{\rL}\PW^{+}_{\rL}$& $  0.61653  ( 5 ) ^{+ 1.0 \%}_{- 1.3 \%}$  &$36.5 $&      $ 0.6799  ( 5 ) ^{+ 0.9 \%}_{- 0.7 \%} $ & $21.9 $ & $         1.103 $ & $1.170$ \\
$\PZ^{\,}_{\rL}\PW^{+}_{\rT}$& $  0.06444  ( 1 ) ^{+ 0.7 \%}_{- 1.0 \%}$  &$3.8 $&       $ 0.17584  ( 6 ) ^{+ 10.8 \%}_{- 8.6 \%} $ & $5.7 $ & $          2.729  $& $1.158 $  \\
$\PZ^{\,}_{\rT}\PW^{+}_{\rL}$& $  0.07437  ( 1 ) ^{+ 0.6 \%}_{- 0.9 \%}$  &$4.4 $&       $ 0.24742  ( 8 ) ^{+ 11.0 \%}_{- 8.9 \%} $ & $8.0 $ & $          3.327  $& $ 1.193 $  \\
$\PZ^{\,}_{\rT}\PW^{+}_{\rT}$& $  0.88233  ( 9 ) ^{+ 2.9 \%}_{- 2.9 \%}$  &$52.3 $&      $ 2.0041  ( 8 ) ^{+ 9.6 \%}_{- 7.7 \%} $ & $64.3 $ & $         2.271 $ & $ 1.227 $ \\
interf. & $ 0.0503(3) $  &$3.0 $&      $0.004(2) $  &$0.1 $ & $-$ & $- $\\
\hline
\end{tabular}
\end{center}
\caption{
  Integrated cross-sections (in fb) in the resolved and unresolved fiducial
  set\-ups described in \refse{subsec:setupscale} for unpolarised
  and doubly-polarised $\PZ\PW^+$~production in the semi-leptonic decay channel.
  Polarisations are defined in the di-boson CM frame.
  Numerical errors (in parentheses) and QCD-scale uncertainties from 7-point
  scale variations (in percentages) are shown.
  The fractions (in percentage) are computed as ratios of polarised
  cross-sections over the unpolarised one.
  $K$-factors are defined as ratios of the NLO QCD cross-sections with
  ($K_{\rm NLO}$) and without ($K_{\rm NLO}^{\rm (no\,g)}$)
  gluon-induced contributions over
  the LO ones.
} 
\label{table:sigmainclNLO}
\end{table*}

Before analysing the QCD corrections, we focus on the LO picture that
is already interesting.  The contribution of the LL
polarisation state is rather large ($\approx 35\%$) besides a
sizeable TT contribution ($\approx 50\%$, dominated by left--right
configurations),  while the contribution of the mixed polarisation
states is at the 10\% level.  The suppression of the mixed modes
results from the transverse momentum cuts on the produced vector
bosons of $200\GeV$ and the unitarity suppression of the corresponding
cross-sections with the square of the energy of the longitudinal
vector bosons \cite{Willenbrock:1987xz,Baur:1994ia}.  At variance with
$\PZ\PZ$~production, the LL signal is not suppressed owing to the
triple-gauge-boson coupling that contributes to LO diagrams. At high
energies, the two longitudinal bosons behave in fact like would-be Goldstone
bosons \cite{Cornwall:1974km,Vayonakis:1976vz} that result from an
$s$-channel $\PW^+$ boson carrying the whole partonic energy (which
exceeds $400\GeV$ in our setups).
While in the resolved topologies the mixed polarisation states have
similar cross-sections, in the unresolved topology the
TL~cross-section 
is $15\%$ larger than the LT one. This is due to the
fact that the two quarks from the decay of a transverse 
$\PW$~boson are preferably produced in and opposite to the direction
of the $\PW$~boson in the $\PW$-boson CM frame.  Thus,
they are roughly back to back in the laboratory frame and less likely
recombined to a fat jet.  Therefore, the two-light-jet requirement is
fulfilled more easily than the requirement of a single fat jet with a
mass close to $\MW$.  The decay jets of a longitudinal $\PW$~boson, on
the other hand, are preferably perpendicular to its direction and
consequently more collinear in the laboratory frame and more likely to be
recombined to a fat jet.

The overall picture at NLO QCD shows a smaller difference between the
results in the two setups, especially in the case of LT and TL
polarisations.  The NLO QCD corrections are very large for
polarisation states with at least one transverse boson, with a size
that is comparable to the one for the LO cross-section. On the
contrary, the corrections are small for the purely longitudinal
state. 
The LL state receives
negative corrections (about $-5\%$) in the resolved topologies, while
in the unresolved setup the corrections become positive ($10\%$).  In
the resolved topology the TT and LT states receive comparable
corrections at the $+100\%$ level, while a different behaviour between
the two is found in the unresolved setup.  The largest corrections in
both setups (about $+200\%$) characterise the TL polarisation state.
In general (both for polarised and unpolarised states), these big
corrections are caused by hard QCD radiation in partonic processes
with initial-state gluons \cite{Rubin:2010xp}, which are enhanced by
the large gluon luminosity in the proton. In fact, omitting the
gluon-induced channels, the $K$-factors are much smaller (see column
labelled $K_{\rm NLO}^{\rm (no\,g)}$ in \refta{table:sigmainclNLO}).

\begin{figure}[tb]
  \centering
  \subfigure[$\PZ^{}_{\rL}\PW^+_{\rT}$\label{subfig:ug1}]{\includegraphics[scale=0.55]{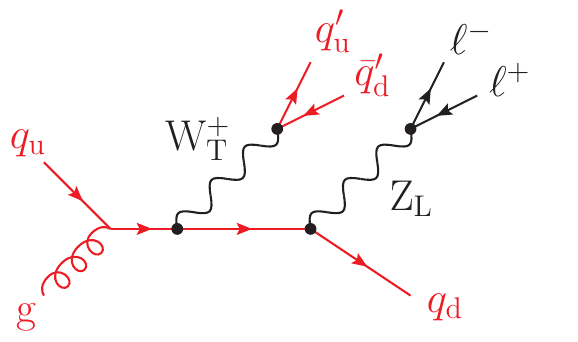}}
  \subfigure[$\PZ^{}_{\rT}\PW^+_{\rL}$\label{subfig:ug2}]{\includegraphics[scale=0.55]{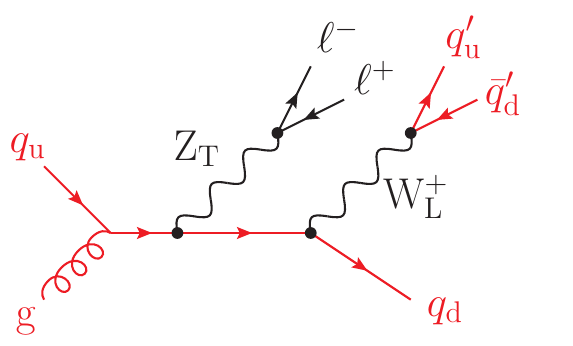}}
  \caption{Leading QCD-radiation contributions in the $q_{\Pu}\Pg$ partonic channel for longitudinal-transverse and transverse-longitudinal $\PZ\PW^+$~production at the LHC in the semi-leptonic decay channel. Particles carrying colour charge are highlighted in red.}\label{fig:ugreal}
\end{figure}

For the LL polarisation state, not only the gluon-emission real
corrections are small but also the gluon-induced ones (with quark
emission) do not give large contributions, as already seen in
inclusive $\PZ\PW$ calculations with leptonic decays
\cite{Denner:2020eck,Le:2022lrp}.  This can be understood as follows.
As demonstrated in \citere{Baglio:2013toa} the dominant NLO QCD
corrections arise for the production of a vector-boson--quark pair in
gluon--quark scattering with subsequent radiation of a soft vector
boson from the quark (see also \reffi{fig:ugreal}). However, the sub-process
$\Pg\Pq\to V\Pq'$ is suppressed for longitudinal high-energy
vector bosons, as can seen via the Goldstone-boson equivalence theorem
and the absence of LO diagrams for the corresponding process with the
vector boson replaced by a would-be Goldstone boson (for massless quarks).

The striking difference (in both setups) between TL and LT at NLO QCD
results from hard QCD radiation in gluon-induced partonic channels which
enhances the mixed state with a longitudinal $\PW$~boson much more
than the one with a transverse $\PW$~boson. This is a consequence of
the unitarity cancellations for mixed polarisation states with high
transverse momenta of the longitudinal vector bosons, which also holds
in the presence of additional real QCD radiation. Owing to this
suppression of high-energy longitudinal bosons, kinematic
configurations are preferred, where the transverse boson recoils
against the longitudinal one and the additional jet, favouring (in the
gluon-induced processes) configurations with a rather collimated
system of the longitudinal boson and the radiation jet.  
We have verified that the Monte Carlo integration channels
corresponding to the  diagrams in \reffis{subfig:ug1} and
\ref{subfig:ug2} give indeed the leading contribution to the LT and TL state,
respectively.
In the TL polarisation mode, all jets are thus produced relatively
close in phase space, resulting in a less efficient decay-jet
identification. In addition, the transverse $\PZ$~boson absorbs the
entire hadronic recoil in the final state, allowing for a softer
longitudinal boson and therefore a less severe unitarity suppression.
For the LT mode the leptonically decaying longitudinal $\PZ$~boson
needs a transverse momentum above $200\GeV$, which causes a stronger
unitarity cancellation.
These effects are further confirmed at differential level, as shown in
\refse{subsec:dif_res}.  Up to these differences, the mixed states
give a contribution of about $14\%$ to the total cross-section at NLO
QCD.

The QCD correction to the TT state is similar in the two setups,
giving a NLO cross-section that represents $65\%$ of the total. It is
worth recalling that the $\PZ\PW$ TT state is characterised by an
approximate amplitude zero at tree level \cite{Baur:1994ia}, which is
spoiled by QCD radiation from higher orders (already at NLO).

The interference effects, obtained subtracting the sum of polarised
cross-sections from the unpolarised one, contribute $+3\%$
at LO and are almost negligible ($<0.5\%$) at NLO QCD,
with irrelevant differences between the two setups. This effect is
more sizeable in differential results.

As expected in general for purely EW processes, the QCD-scale
uncertainties at LO are small since they only come from
factorisation-scale variations.  In addition, the requirement 
of the boosted regime
reduces them roughly by a factor of 4 w.r.t.\ inclusive
calculations \cite{Denner:2020eck,Le:2022lrp}. We have checked
numerically that the same effect is also found in the fully-leptonic
channel for the $\PW\PZ$ process.
At NLO QCD the renormalisation-scale dependence of the strong coupling
and the sizeable real corrections render the scale uncertainties much
larger than at LO, ranging between $7\%$ and $11\%$ for polarisation
modes with at least one transverse boson.  A different behaviour is
found for the LL mode, for which the real corrections are small and
therefore the NLO QCD scale uncertainty is of the same order of
magnitude as at LO ($1\%$).  The difference in scale uncertainties between
the two setups is motivated by the small differences in the applied
kinematic selections.
Owing to the EW character
of the LO $\PZ\PW$ process, a truly NLO QCD scale dependence of the
polarised cross-section can only be obtained upon including NNLO QCD corrections,
which is possible with the tools developed in
\citere{Poncelet:2021jmj}. Notice that this would also provide more
reliable scale uncertainties.

The obtained results (in both setups) understand a hard cut on the transverse momentum of
both bosons, which acts like a veto on additional QCD jets. Although this prevents
huge QCD $K$-factors, the symmetric character of such selections
($\pt{\Pj\Pj/\rm J}>200\GeV,\,\pt{\ell\ell}>200\GeV$) induces large
unphysical NLO corrections in kinematic
regions close to the cut \cite{Salam:2021tbm}. 
These can be avoided by omitting the cut $\pt{\Pj\Pj/\rm
  J}>200\GeV$. The corresponding results for the resolved setup described in
\refse{subsec:setupscale} without this cut are shown in 
\refta{table:loosesetup}.
\begin{table*}[tb]
\begin{center}
\renewcommand{\arraystretch}{1.3}
\begin{tabular}{C{1.7cm}C{2.3cm}C{1.6cm}C{2.3cm}C{1.6cm}C{1.3cm}C{1.3cm}}%
\hline %
\cellcolor{blue!14} state  & \cellcolor{blue!14}  $\sigma_{\rm LO}$ [fb]  & \cellcolor{blue!14} $f_{\rm LO}[\%]$& \cellcolor{blue!14} $\sigma_{\rm NLO}$ [fb] & \cellcolor{blue!14} f$_{\rm NLO}[\%]$ & \cellcolor{blue!14} {$K_{\rm NLO}$}   & \cellcolor{blue!14} {$K_{\rm NLO}^{\rm (no\,g)}$}    \\
\hline
\multicolumn{7}{l}{\cellcolor{green!9} resolved (no minimum $\pt{\Pj\Pj}$ cut), $\PZ(\Pe^+\Pe^-)\PW^+(\Pj\Pj)$} \\
\hline
unpol. &         $  1.8564 ( 1 )   ^{+ 1.2 \%}_{- 1.4 \%}$  & $100 $ &   $ 5.5388  ( 8 ) ^{+ 10.6 \%}_{- 8.6 \%} $   & $100$ & $          2.984 $  & $ 1.371 $ \\
$\PZ^{\,}_{\rL}\PW^{+}_{\rL}$  &         $  0.64605  ( 3 ) ^{+ 0.2 \%}_{- 0.6 \%}$  & $34.8 $ &          $ 0.7525  ( 4 ) ^{+ 1.5 \%}_{- 1.2 \%} $ &$13.6$ & $     1.165 $ & $ 1.194 $ \\
$\PZ^{\,}_{\rL}\PW^{+}_{\rT}$  &         $  0.08687  ( 1 ) ^{+ 0.2 \%}_{- 0.6 \%}$  & $4.7 $ &   $ 0.3057  ( 1 ) ^{+ 11.4 \%}_{- 9.2 \%} $ & $5.5$ & $    3.519 $ & $ 1.462 $ \\
$\PZ^{\,}_{\rT}\PW^{+}_{\rL}$  &         $  0.08710  ( 1 ) ^{+ 0.1 \%}_{- 0.6 \%}$  & $4.7 $ &   $ 1.0486  ( 1 ) ^{+ 14.6 \%}_{- 11.9 \%} $ & $18.9$ & $    12.04 $ & $2.408  $ \\
$\PZ^{\,}_{\rT}\PW^{+}_{\rT}$  &         $  0.97677  ( 7 ) ^{+ 2.0 \%}_{- 2.2 \%}$  & $52.6 $ &          $ 3.5506  ( 9 ) ^{+ 11.8 \%}_{- 9.6 \%} $ & $64.1$& $    3.635 $ & $ 1.424 $ \\
interf.  & $ 0.0595(1) $  &$3.2 $&     $-0.119(2)$  &$ -2.1 $ & $-$ & $ - $ \\
\hline
\end{tabular}
\end{center}
\caption{
  Integrated cross-sections (in fb) in the resolved
  setup described in \refse{subsec:setupscale} without the minimum $\pt{\Pj\Pj}$ cut of $200\GeV$. 
  Polarisations are defined in the di-boson CM frame.
  Numerical errors (in parentheses) and QCD-scale uncertainties from 7-point
  scale variations (in percentages) are shown.
  The fractions (in percentage) are computed as ratios of polarised
  cross-sections over the unpolarised one.
  $K$-factors are defined as ratios of the NLO QCD cross-sections
  with ($K_{\rm NLO}$) and without ($K_{\rm NLO}^{\rm (no\,g)}$) gluon-induced contributions
  over the LO ones.
} 
\label{table:loosesetup}
\end{table*}
While the LO picture is identical to the original setup, the omission
of the $\pt{\Pj\Pj/\rm J}$ cut induces huge QCD corrections owing
to the fact that the large transverse momentum of the leptonic system
is now absorbed by the entire hadronic system, including both decay
and extra real radiation jets. This leads to much larger
$K$-factors for all polarisation states with at least one transverse
boson, while the LL contribution does not change so much, confirming
the arguments given above.
The QCD corrections and, in
particular, the gluon-induced contributions are huge for the TL state,
giving a TL component that amounts at almost $19\%$ of the total
unpolarised result (compared to $5\%$ at LO).  As expected, the TL
contribution is much larger than the LT one at NLO QCD, because the
$\PZ$~boson is boosted ($\pt{\ell\ell}>200\GeV$) while the
hadronically decaying $\PW$~boson is not necessarily boosted,
therefore the unitarity suppression is strong for the LT state but
much smaller for the TL state. Note that the strong increase of the
mixed polarisation states is again dominantly due to the gluon-induced channels,
as can be seen by inspecting the $K$-factors $K_{\rm NLO}^{\rm
  (no\,g)}$ without these contributions in \refta{table:loosesetup}.
Despite the small QCD corrections, the LL fraction is diminished
($13.6\%$) w.r.t.\ the default setups ($20\%$), but remains sizeable
compared to the $6\%$ found in inclusive fiducial setups
\cite{Denner:2020eck,Le:2022lrp}.

  We summarise the main results for fiducial $\PZ\PW^+$ cross-sections in
  the boosted setups considered in this paper. With about $35\%$ of the total,
  the LL polarisation state gives a pretty large contribution at LO to the $\PZ\PW^+$
  process, while it gets suppressed to about $20\%$ when including NLO QCD corrections.
  This is related to the fact that the NLO QCD corrections are small
  for the LL final state, while they are large if at least one transverse vector boson is present.
  The corrections are largest in the case of one longitudinal $\PW$~boson and a
  transverse $\PZ$~boson. The impact of interferences between longitudinal and
  transverse polarisations is below $0.5\%$ for the NLO QCD results.
  Omitting the cut $\pt{\Pj\Pj/\rm J}>200\GeV$, the NLO QCD corrections to polarised
  signals with transverse $\PZ$~bosons are strongly enhanced.
  
  As a last comment of this section, we report on a LO study we performed
  to give a rough estimate of the most relevant irreducible backgrounds
  to resonant $\PZ\PW^+$ production in the $\Pe^+\Pe^-\Pj\Pj$ final state.
  Two more resonant structures contribute, namely $\PZ\PW^-$ and
  $\PZ\PZ$ production,
  which have been computed with the DPA technique described in
  \refse{sec:outline_subsec:DPA}. The sum of the $\PZ\PW^+$,
  $\PZ\PW^-$, and $\PZ\PZ$ resonant processes gives the dominant contribution to the full
  off-shell process at order $\mc O(\alpha^4)$, up to non-resonant effects
  beyond the DPA. The same final state also receives contributions of QCD type,
  formally of order $\mc O(\alphas^2\alpha^2)$ (pure QCD background) and $\mc O(\alphas\alpha^3)$
  (QCD--EW interference), involving a $\PZ$ boson decaying leptonically and
  two QCD-originated jets.
  The LO cross-sections for such backgrounds are shown in Table~\ref{table:comparison_background},
  for both setups introduced in \refse{subsec:setupscale}.
  The resonant $\PZ\PW^-$ and $\PZ\PZ$ processes turn out to be roughly of the same size
  as the unpolarised $\PZ\PW^+$ signal, with a $\PZ\PZ$ contribution that is even larger
  than $\PZ\PW^+$ in the resolved setup. This hierarchy is inverted in the unresolved
  setup, thanks to kinematic selections that cut away more effectively the $\PZ\PZ$ background.
  The effects beyond the DPA, estimated as the difference between the full off-shell EW
  result and the sum of the three DPA results, are at the 2--4\% level, as expected
  from the formal DPA accuracy of $\mc O(\Gamma_V/M_V)$.
  The pure-QCD non-resonant background has a severe impact on the cross-section,
  being 20(15)-times larger than the EW result in the resolved(unresolved) setup. The
  interference contribution is negative but much smaller than the pure-QCD one,
  with a size about $6\%$ of the full EW cross-section.
  As already mentioned in \refse{sec:outline}, in polarisation
  analyses it is crucial to
  keep under control and possibly subtract all non-resonant backgrounds, especially if
  their size is larger than the targeted signal, \eg QCD multi-jet production.
  We stress again that, although they are part of the same full off-shell EW signal, the three
  di-boson processes $\PZ\PW^+$, $\PZ\PW^-$ and $\PZ\PZ$ are characterised by very different
  resonance and spin structures, with LO suppressions in different polarisation states and/or in different kinematic regions.
  Therefore, including the three of them in the same polarisation analysis
  could end up in washing out spin-specific details in differential distributions and
  covering new-physics effects affecting a specific resonant process with spurious
  effects originated from other SM processes.
  In the light of extracting doubly-polarised $\PZ\PW^+$ production from LHC data
  in the semileptonic channel, it is then crucial to subtract all of the mentioned backgrounds
  before further splitting the unpolarised $\PZ\PW^+$ signal into
  polarised sub-signals.

\begin{table*}
  \begin{center}
    \renewcommand{\arraystretch}{1.3}
      \begin{tabular}{c|cc|cc}
        \hline %
        \cellcolor{green!9} & \multicolumn{2}{c|}{\cellcolor{green!9} resolved} & \multicolumn{2}{c}{\cellcolor{green!9} unresolved}\\
        \hline
        \cellcolor{blue!14} {process}
        & \cellcolor{blue!14}  {$\sigma_{\rm LO}$ [fb]}
        & \cellcolor{blue!14} {ratio over full $O(\alpha^4)$}
        & \cellcolor{blue!14}  {$\sigma_{\rm LO}$ [fb]}
        & \cellcolor{blue!14} {ratio over full $O(\alpha^4)$}\\
        \hline
        DPA $\PZ\PW^+$ & $1.8567(2)_{- 1.4 \%}^{+ 1.2 \%}$ & $0.353$ & $1.6879(2)_{- 2.1 \%}^{+ 1.9 \%}$ & $0.425$ \\
        DPA $\PZ\PW^-$ & $1.0527(1)_{- 1.6 \%}^{+ 1.3 \%}$ & $0.200$ & $0.9003(1)_{- 2.1 \%}^{+ 2.0 \%}$ & $0.227$ \\
        DPA $\PZ\PZ^{\quad}$ & $2.1430(3)_{- 1.6 \%}^{+ 1.3 \%}$ & $0.408$ & $1.2804(2)_{- 2.7 \%}^{+ 2.6 \%}$ & $0.323$ \\
        \,DPA $\PZ V^{\quad}$ & $5.0523(4)_{- 1.5 \%}^{+ 1.3 \%}$ & $0.961$ & $3.8685(3)_{- 2.3 \%}^{+ 2.2 \%}$ & $0.975$ \\
        full $O(\alpha^4)\quad$ & $5.253(1)_{- 1.5 \%}^{+ 1.2 \%}$ & $1.000$ & $3.967(2)_{- 2.3 \%}^{+ 2.1 \%}$ & $1.000$ \\
        full $O(\alphas\alpha^3)$ & $-0.3124(6)_{- 10.7 \%}^{+ 9.2 \%}$ & $-0.059$ & $-0.2145(6)_{- 11.4 \%}^{+ 9.7 \%}$ & $-0.054$ \\
        full $O(\alphas^2\alpha^2)$ & $97.91(7)_{- 18.4 \%}^{+ 24.3 \%}$ & $18.638$ & $62.55(7)_{- 18.8 \%}^{+ 25.0 \%}$ & $15.768$ \\
        \hline
      \end{tabular}
  \end{center}
  \caption{
      Integrated LO cross-sections (in fb) in the resolved and unresolved fiducial
      set\-ups described in \refse{subsec:setupscale} for:
      the signal process (DPA $\PZ\PW^+$), the background processes
      with a resonant $\PW^-$ boson (DPA $\PZ\PW^-$) and with two resonant $\PZ$~bosons (DPA $\PZ\PZ$),
      the full off-shell process at the three perturbative orders in $\alphas$.
      The result dubbed DPA $\PZ V$ is understood as the sum of the three DPA contributions
      ($ \PZ\PW^+ + \PZ\PW^- + \PZ\PZ$).
  }
  \label{table:comparison_background}
\end{table*}

\subsection{Differential Results}
\label{subsec:dif_res}
In order to understand the relative importance of the various polarisation states
and the differences between the two setups at LO and at NLO QCD, it is essential to
analyse differential distributions, which are presented in this section.
Unless otherwise stated, the two default setups described in
\refse{subsec:setupscale} are understood. 

\subsubsection{Distributions in the decay angles and the scattering angle}

\begin{figure}[tb]
  \centering
  \subfigure[Resolved topology \label{fig:dec_res}]  {\includegraphics[scale=0.39]{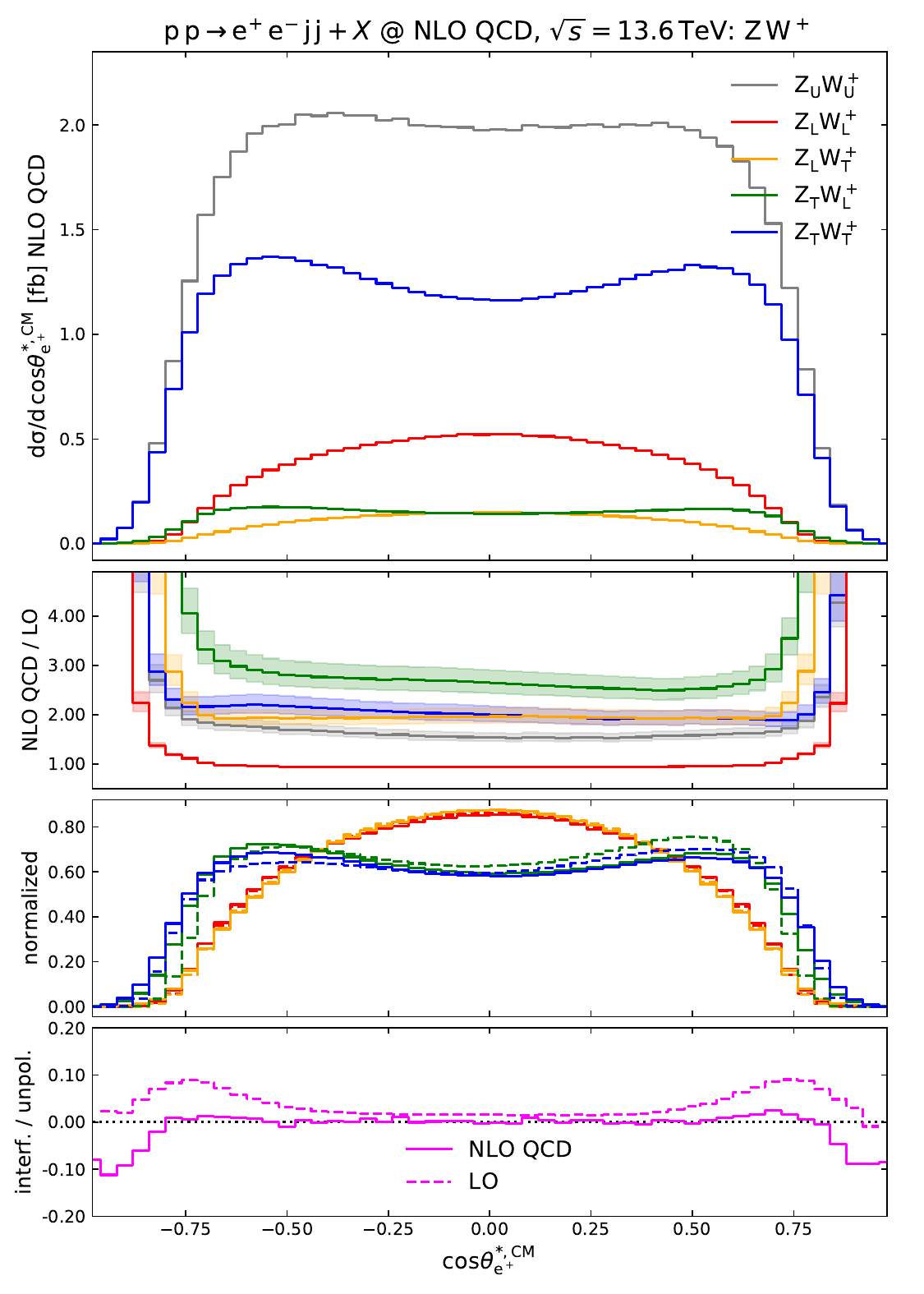}}
  \subfigure[Unresolved topology \label{fig:dec_unres}]{\includegraphics[scale=0.39]{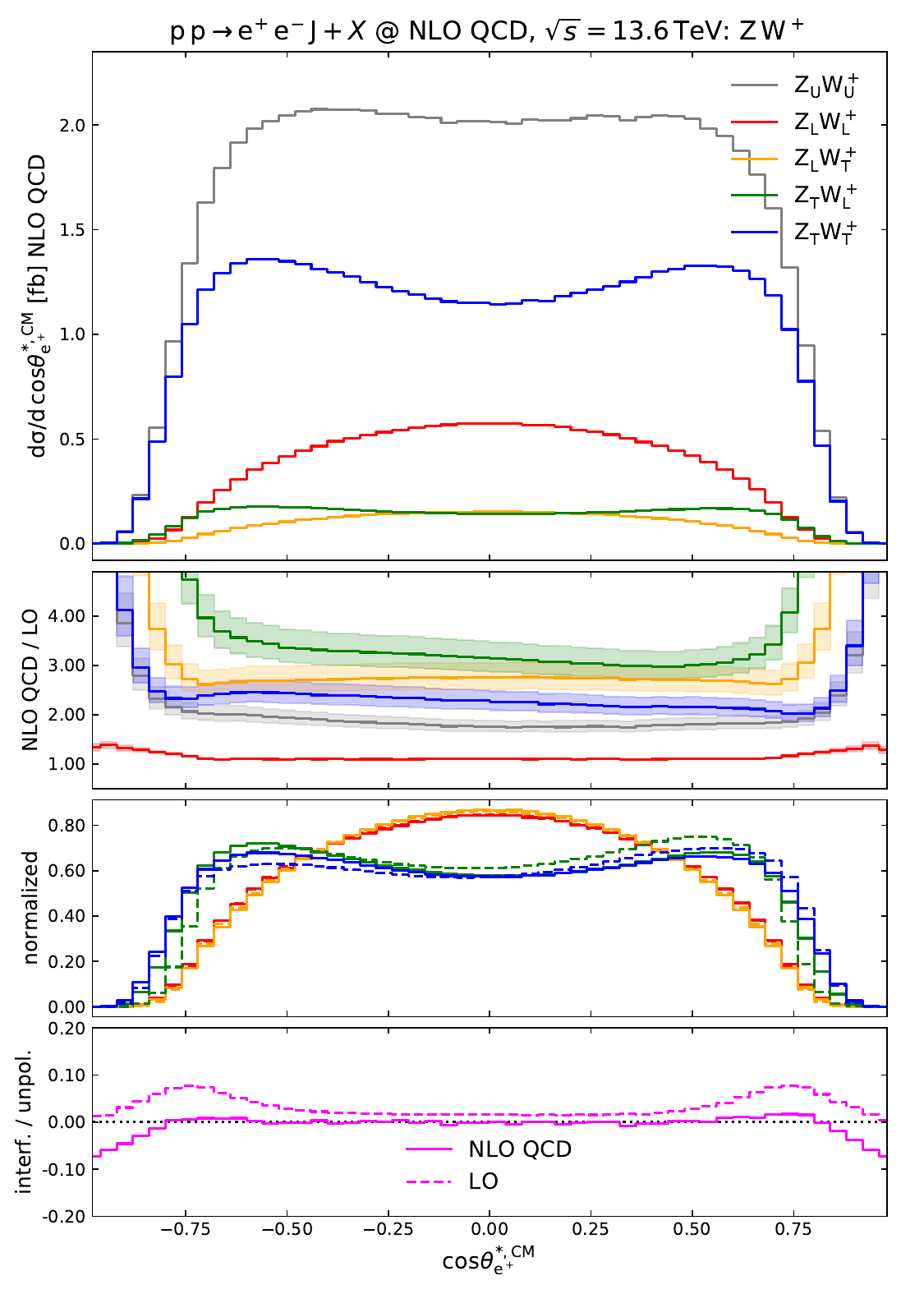}}
  \caption{
    Distribution in the cosine of the polar decay angle of the
    positron in semi-leptonic $\PZ\PW^+$~production at the LHC.
    The definition of this angle is given in Eq.~\eqref{eq:thetastar}.
    Results for the unpolarised and doubly-polarised
    process are shown in the resolved (left) and unresolved
    (right) setups described in \refse{subsec:setupscale}.
    From top down: NLO QCD differential cross-sections,
    ratios of NLO QCD cross-sections over the LO ones,
    normalised LO (dashed) and NLO QCD (solid) shapes (unit integral),
    interference contributions relative to the unpolarised
    cross-section.    
  }\label{fig:dec} 
\end{figure}

We start the discussion on differential results with the 
angular observable that is directly related to the polarisation state
of an EW boson.  The polar decay angle of the lepton $\Pe^\pm$
is defined as the angular separation between the lepton
direction in the rest frame of the leptonically-decaying boson
($\vec{p}^{\,*}_{\Pe^\pm}$) and the direction of the same boson
calculated in the reconstructed di-boson CM frame
($\vec{p}^{\:\CM}_{\Pe^+\!\Pe^-}$),
\beq\label{eq:thetastar}
\cos\theta^{*,\rm CM}_{\Pe^\pm}\,=\,\frac{\vec{p}^{\,*}_{\Pe^\pm} \cdot \vec{p}^{\:\CM}_{\Pe^+\!\Pe^-}}{|\vec{p}^{\,*}_{\Pe^\pm}||\vec{p}^{\:\CM}_{\Pe^+\!\Pe^-}|}\,.
\eeq
The differential cross-section with respect to $\cos\theta^{*,
  \text{CM}}_{\Pe^+}$ is depicted in \reffi{fig:dec}.
The polar decay angle $\theta^{*,\rm CM}_{\Pe^+}$ is very well suited for the discrimination
between polarisations states of the $\PZ$~boson, while it cannot give
access to the polarisation state of the other boson, therefore very
similar shapes are expected for LL and LT modes, as well as for the TL
and TT ones. The slight differences are a consequence of the (small)
interference and spin-correlation effects.  As expected from results
in the fully-leptonic decay channel \cite{Denner:2020eck,Le:2022lrp},
a longitudinally polarised $\PZ$~boson gives leptons produced mostly
around $\theta^{*,\rm CM}_{\Pe^+}=\pi/2$, while a transverse one populates more regions
around $\cos \theta^{*,\rm CM}_{\Pe^+}=\pm 0.6$. In both cases the collinear and
anti-collinear configurations are suppressed by the transverse-momentum
cut on the charged leptons. The NLO QCD corrections
distort the polarised and unpolarised shapes only in less-populated
regions. In the rest of the spectrum they give rather flat
enhancements to the LL and LT distributions, while for the TT and TL
states the $K$-factors mildly diminish towards positive
$\cos\theta^{*,\rm CM}_{\Pe^+}$. In the regions where the TT and TL distributions
feature a sharp drop, the interferences reach the $9\%$ level at LO,
while they almost disappear at NLO QCD.  Overall, the leptonic decay
of the $\PZ$~boson is only indirectly affected by QCD effects.  The
results for this observables are basically identical in the two
setups.

\begin{figure}[tb]
  \centering
  \subfigure[Resolved topology \label{fig:scatt_res}]  {\includegraphics[scale=0.39]{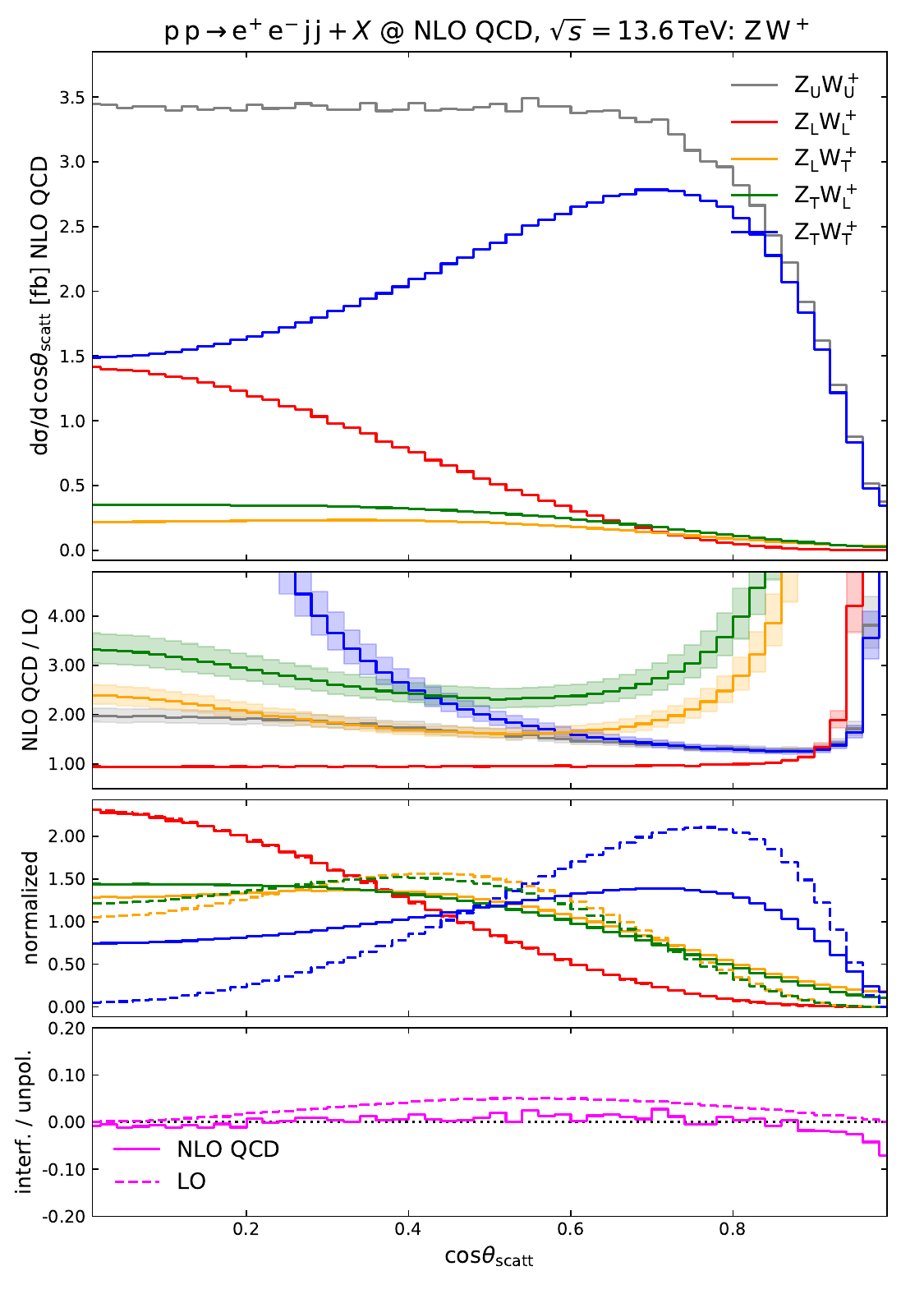}}
  \subfigure[Unresolved topology \label{fig:scatt_unres}]{\includegraphics[scale=0.39]{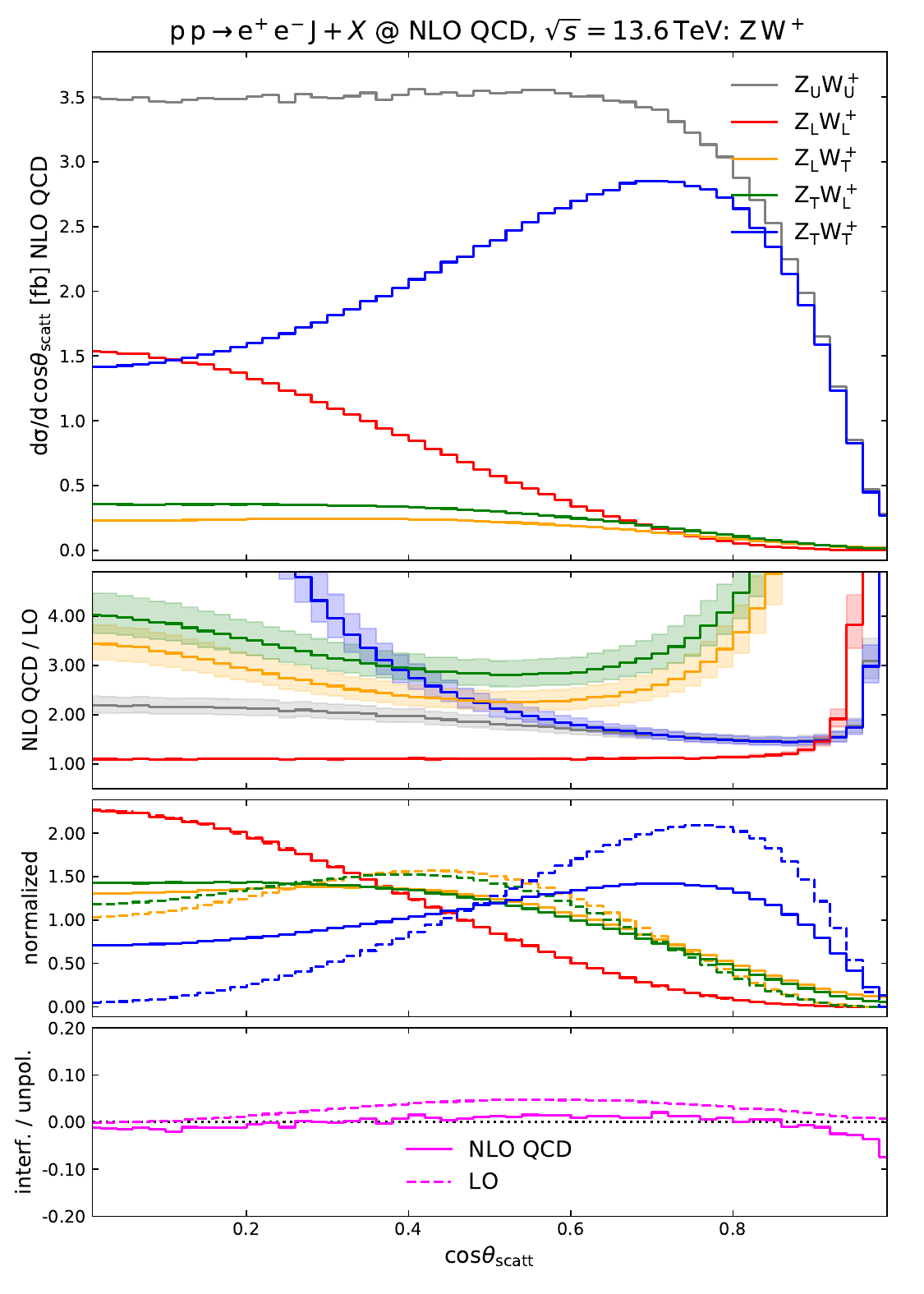}}
  \caption{
    Distribution in the cosine of the scattering angle
    in semi-leptonic $\PZ\PW^+$~production at the LHC.
    The scattering angle is defined according to Eq.~\eqref{eq:thetascatt}.
    Same structure as \reffi{fig:dec}.
  }\label{fig:scatt} 
\end{figure}

In \reffi{fig:scatt} the differential cross-section in the cosine of the scattering angle is shown.
This angle is defined as
\beq\label{eq:thetascatt}
\cos\theta_{\rm scatt}\,=\,\frac{\big|p^{\CM}_{\Pe^+\Pe^-,\,\rm z}\big|}{\big|{\vec{\,p}}^{\,\,\CM}_{\Pe^+\Pe^-}\big|}\,,
\eeq
where ${\vec{\,p}}^{\,\,\CM}_{\Pe^+\Pe^-}$ is the three momentum of the electron--positron pair in the reconstructed di-boson CM frame
and $p^{\CM}_{\Pe^+\Pe^-,\,\rm z}$ is its component in the $\hat{z}$ direction.
This observable is directly related to the  production level, while it is expected to be rather decay agnostic, up to
small effects due to the decay-product reconstruction. 
In \reffi{fig:scatt}, manifest differences can be seen for various polarisation modes. 
At large $\cos\theta_{\rm scatt}$ the cross-section is dominated by the TT polarised state, while
around zero the LL and the TT states give almost the same contribution to the unpolarised result.
The QCD corrections sizeably distort the shapes of the LT, TL and TT
distributions, with a particularly significant effect in the TT case.
The marked shape change for the TT state is determined by the steeply
increasing $K$-factor at low $\cos\theta_{\rm scatt}$. This is caused
by real corrections spoiling the approximate amplitude-zero effect
which is present at LO in the quark--antiquark annihilation into
$\PW\PZ$ \cite{Baur:1994ia,Liu:2018pkg}.  Very large
$K$-factors are also found for the LT and TL polarisation modes in
forward-scattering regions, where the LO signals are suppressed by
unitarity cancellations due to
the large transverse-momentum cuts applied on the two bosons,
requiring very high scattering energy when combined with
forward/backward scattering angles.  All $K$-factors around
$\cos\theta_{\rm scatt}=1$ become huge, because real radiation leaves
room to configurations where the boson trajectory is closer to the
beam direction, without being cut away by the rapidity selections or suppressed.
Interferences at NLO QCD are almost vanishing in the most populated
region, while they become non-negligible in forward-scattering
configurations.  The discrimination power of this angular variable
amongst the polarisation states is marked. However, using this
observable is well motivated only for SM studies and measurements,
since the model independence of the polarised shapes is not given.
In fact, at variance with decay angles, the
distribution shapes in the scattering angle could vary sizeably
depending on the production dynamics, \ie it is very sensitive to
new-physics effects at production level.

\begin{figure}[tb]
  \centering
  \subfigure[Resolved topology \label{fig:yj_res}]  {\includegraphics[scale=0.39]{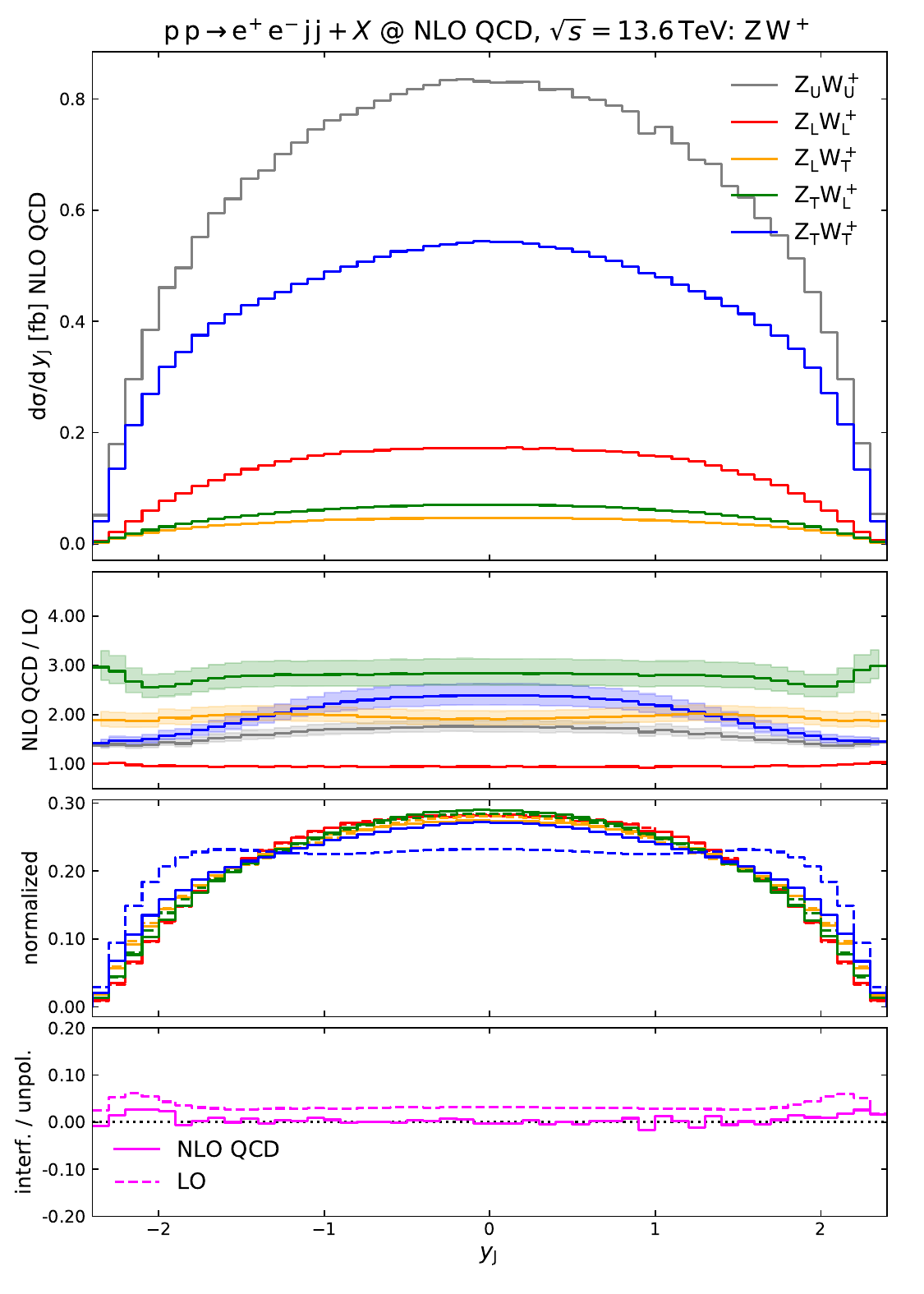}}
  \subfigure[Unresolved topology \label{fig:yj_unres}]{\includegraphics[scale=0.39]{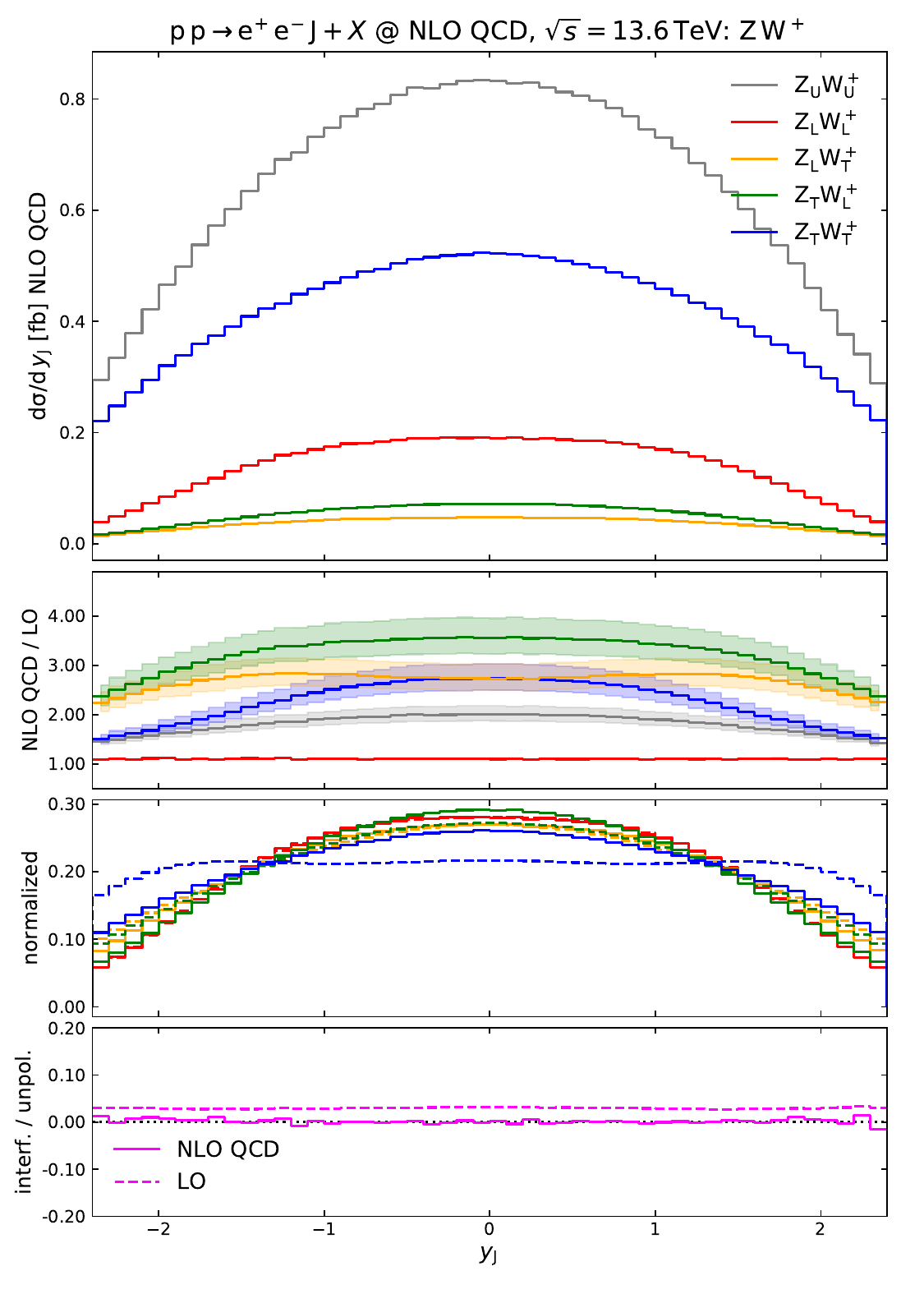}}
  \caption{
    Rapidity distribution of the hadronic system $\rm J$  
    in semi-leptonic $\PZ\PW^+$~production at the LHC.
    The identification of the hadronic system  $\rm J$ is described in \refse{subsec:setupscale}.
    Same structure as \reffi{fig:dec}.
  }\label{fig:yj} 
\end{figure}

{\samepage
\subsubsection{Rapidity distributions}\nopagebreak
%
In \reffi{fig:yj} the differential cross-section in the rapidity of
the hadronic system~$\mathrm{J}$ [defined after Eq.~\refeq{eq:mtdef}]
is considered.}
In the case where both bosons are transversely polarised the NLO QCD
corrections significantly change the shape of the distribution.  This
is caused by the large contributions from real emission via
gluon-induced processes, which mostly fill the central region, while
at LO the distributions is almost flat for $|y_{\rm J}|<1.8$. This is
an indirect effect of the approximate amplitude zero at LO, as already observed in
\reffi{fig:scatt} for the distributions in the scattering angle.  In
the resolved setup, large flat corrections are found for the TL and LT
contributions, with mild non-flat effects just around $|y_{\rm J}|=2$.
In the unresolved setup, the mixed states are characterised by less
flat $K$-factors, giving slightly more sizable shape changes than in
the resolved case.  In both setups, the LL distribution only receives
small corrections reflecting the result at integrated level.  The
interferences at NLO QCD are very small in the whole accessible
spectrum, and negligible in the unresolved setup.  Overall,
the different shapes in the two setups are due to the sharp cut
$|y_{\rm J}|<2.4$ in unresolved topologies, which is replaced in the
resolved ones by rapidity cuts on single jets that suppress the region
$|y_{\Pj\Pj}|\gtrsim 2.4$.  This causes the steeper fall off at the
edges of the distribution of the resolved setup compared to the
unresolved setup.

\begin{figure}[tb]
  \centering
  \subfigure[Resolved topology \label{fig:yz_res}]  {\includegraphics[scale=0.39]{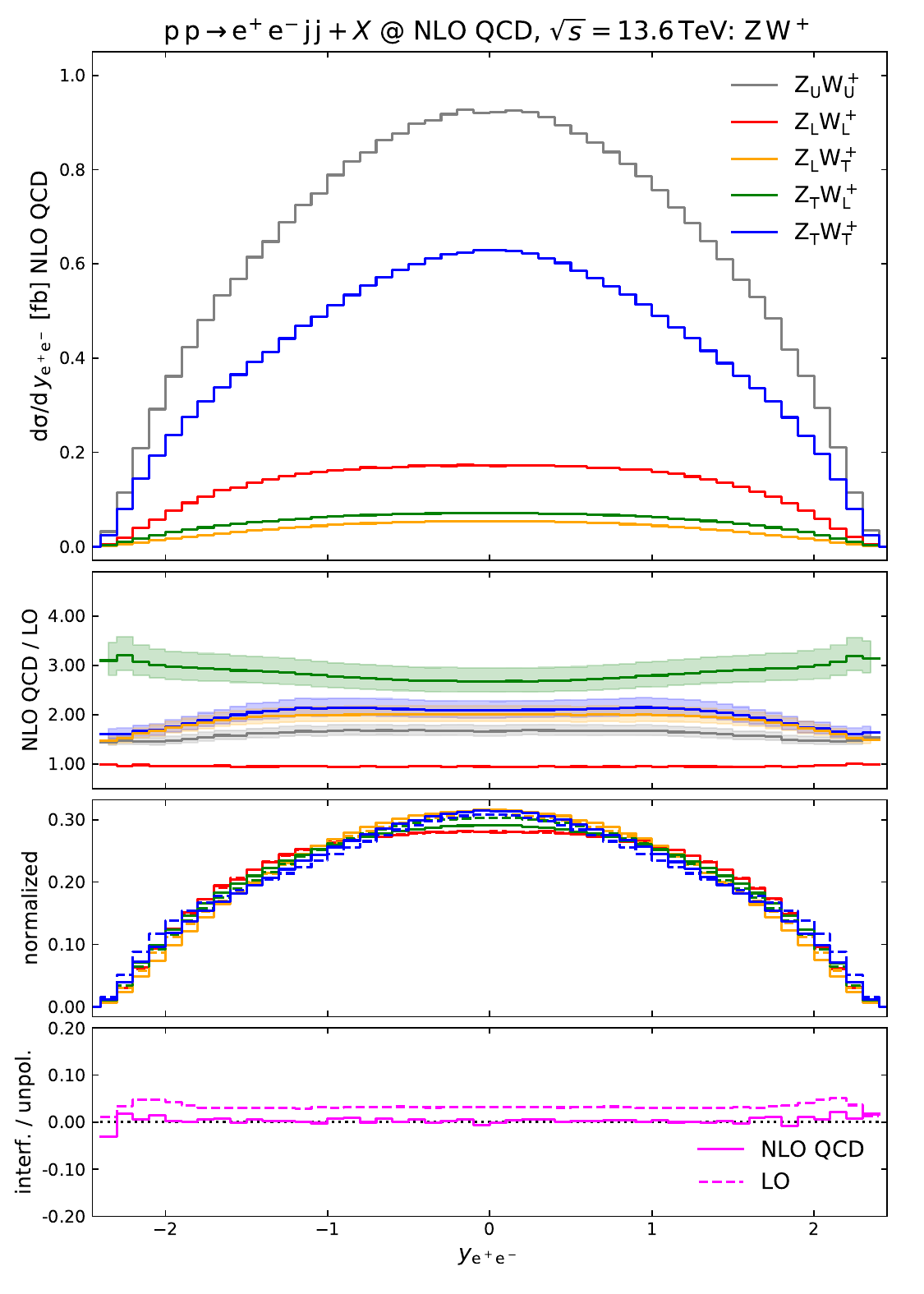}}
  \subfigure[Unresolved topology \label{fig:yz_unres}]{\includegraphics[scale=0.39]{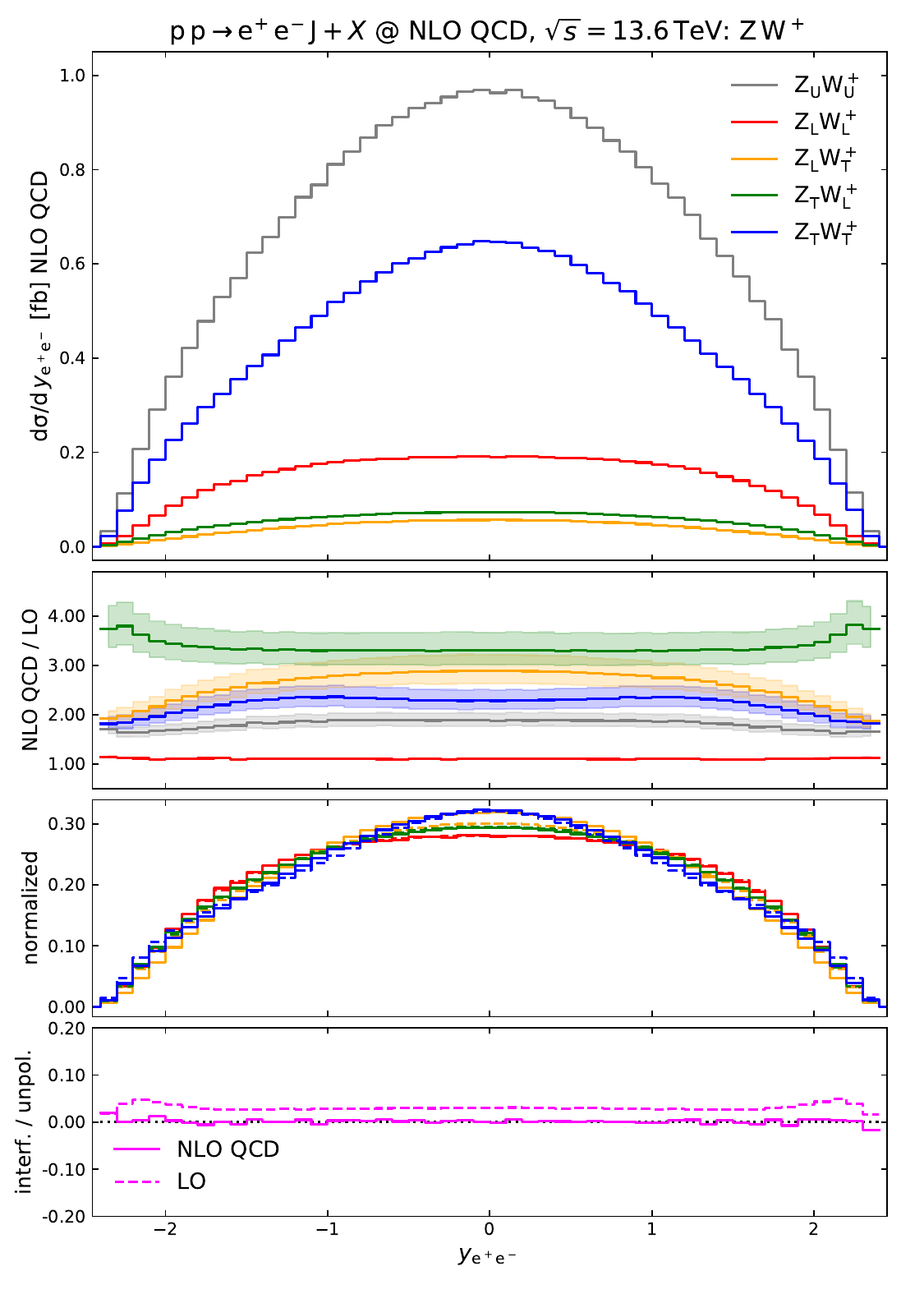}}
  \caption{
    Rapidity distribution of the electron--positron pair  
    in semi-leptonic $\PZ\PW^+$~production at the LHC.
    Same structure as \reffi{fig:dec}.
  }\label{fig:yz} 
\end{figure}

In \reffi{fig:yz} the distribution in the rapidity of the electron--positron pair is presented. 
This observable is strongly correlated to the rapidity of the hadronic
system~$\rm J$ discussed
in \reffi{fig:yj}. In fact, the $\PZ$-boson momentum absorbs the recoil of the entire hadronic
system, including the $\PW$~boson and additional QCD radiation. 
Very small shape modifications are found comparing LO and NLO QCD distributions.
In fact, a non-flat behaviour of the $K$-factors
is only found in the largest-rapidity regions that are not forbidden by selection cuts,
with a mild increase towards $|y_{\Pe^+\Pe^-}|=2.4$ for the TL state, a mild decrease in the same
region for the LT and TT states. Up to different overall normalisations, the corrections
behave very similarly in the two setups.
The shapes for different polarisation states are very close to each other.
Only in the central region the TT component is slightly more peaked than the others. 
Interference contributions are basically independent of $y_{\Pe^+\Pe^-}$.

\begin{figure}[tb]
  \centering
  \subfigure[Resolved topology \label{fig:dyej_res}]  {\includegraphics[scale=0.39]{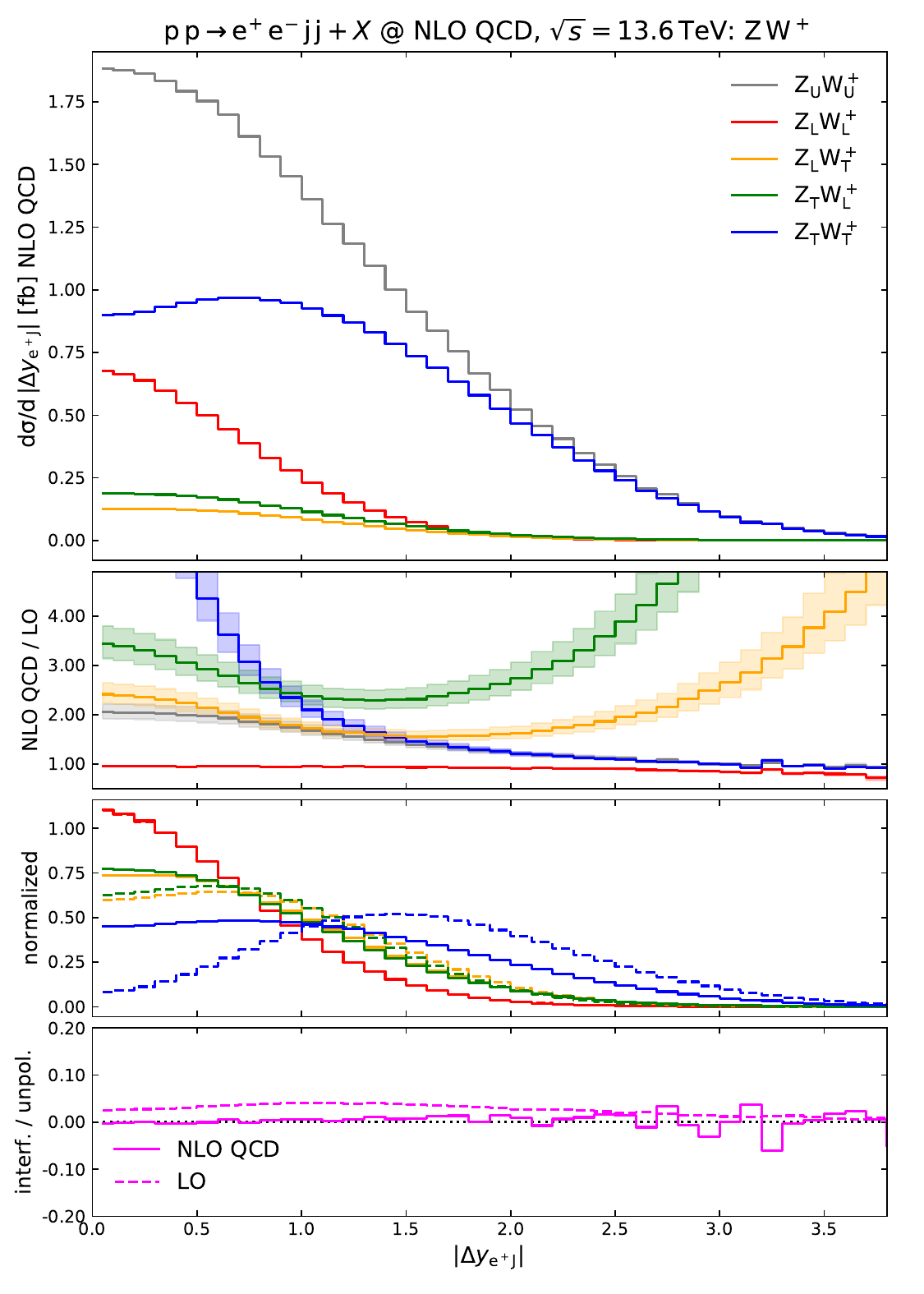}}
  \subfigure[Unresolved topology \label{fig:dyej_unres}]{\includegraphics[scale=0.39]{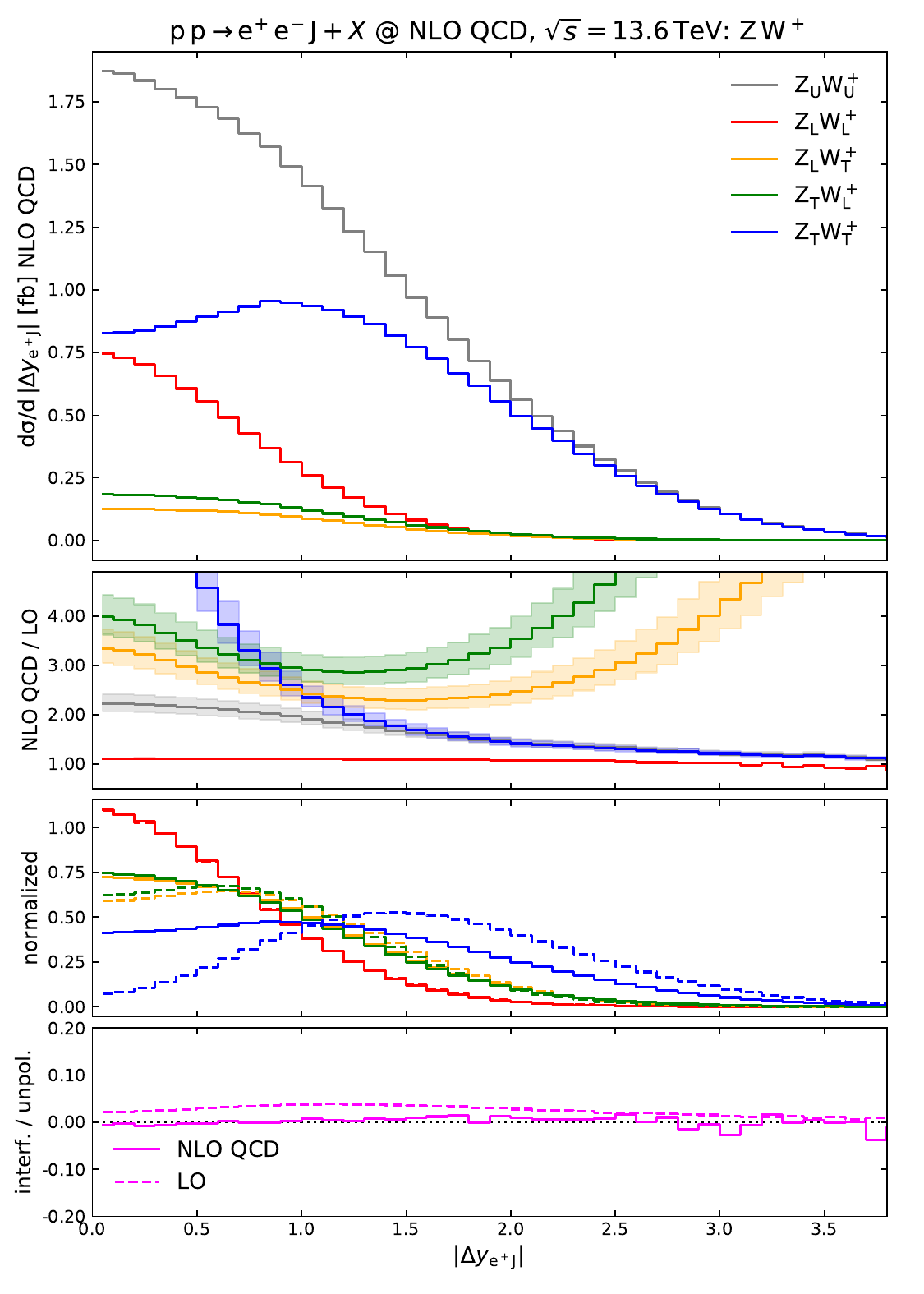}}
  \caption{
    Distribution in the rapidity separation between the
    positron and the hadronic system $\rm J$
    in semi-leptonic $\PZ\PW^+$~production at the LHC.
    The identification of the hadronic system  $\rm J$
    is described in \refse{subsec:setupscale}.
    Same structure as \reffi{fig:dec}.
  }\label{fig:dyej} 
\end{figure}

The distribution shown in \reffi{fig:dyej} concerns the absolute value of the rapidity
difference between the positron and the hadronic system  $\rm J$. 
Similarly to the scattering angle, this observable is well suited to
discriminate amongst different polarisation states, thanks to the
marked shape differences.  Concerning model dependence, the same
caveats apply as for the distribution in the scattering angle.
The LL, the TL, and the LT polarisation
states have a maximum at $|\Delta y_{\Pe^+\rm J}| = 0$, while the
maximum of the TT state is shifted to $|\Delta y_{\Pe^+\rm J}| \approx
0.65$. The LO TT shape is heavily distorted by QCD corrections, again
due to real radiation that fills kinematic regions that are suppressed
at LO due to the approximate amplitude zero ($|\Delta y_{\Pe^+\rm J}| \approx 0$).
The LT and TL polarisation states receive large and increasing
corrections from real radiation in gluon-induced channels for $|\Delta
y_{\Pe^+\rm J}| \gtrsim 2.5$, where the LO is extremely suppressed.
In general, at large rapidity separation all signals are suppressed
by the rapidity cuts.  Interferences are very small through the
whole spectrum at NLO QCD.
Apart from different total cross-sections, the two kinematic setups
give almost identical results for all polarisation states, both in
terms of normalised shapes and in terms of $K$-factor behaviours.\\

\subsubsection{Invariant-mass distributions}

\begin{figure}
  \centering
  \subfigure[Resolved topology \label{fig:mj_res}]    {\includegraphics[scale=0.39]{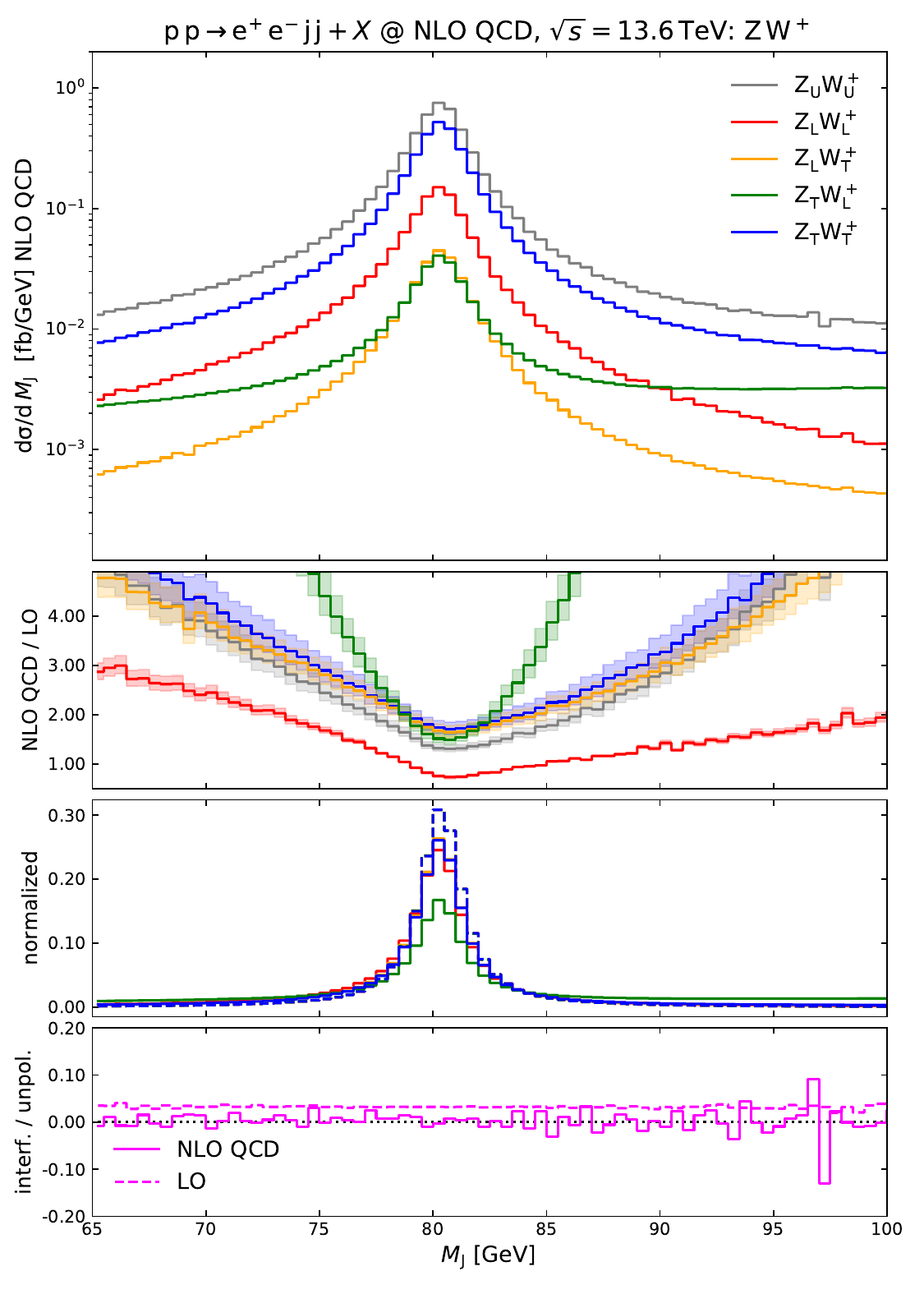}}
  \subfigure[Unresolved topology \label{fig:mj_unres}]{\includegraphics[scale=0.39]{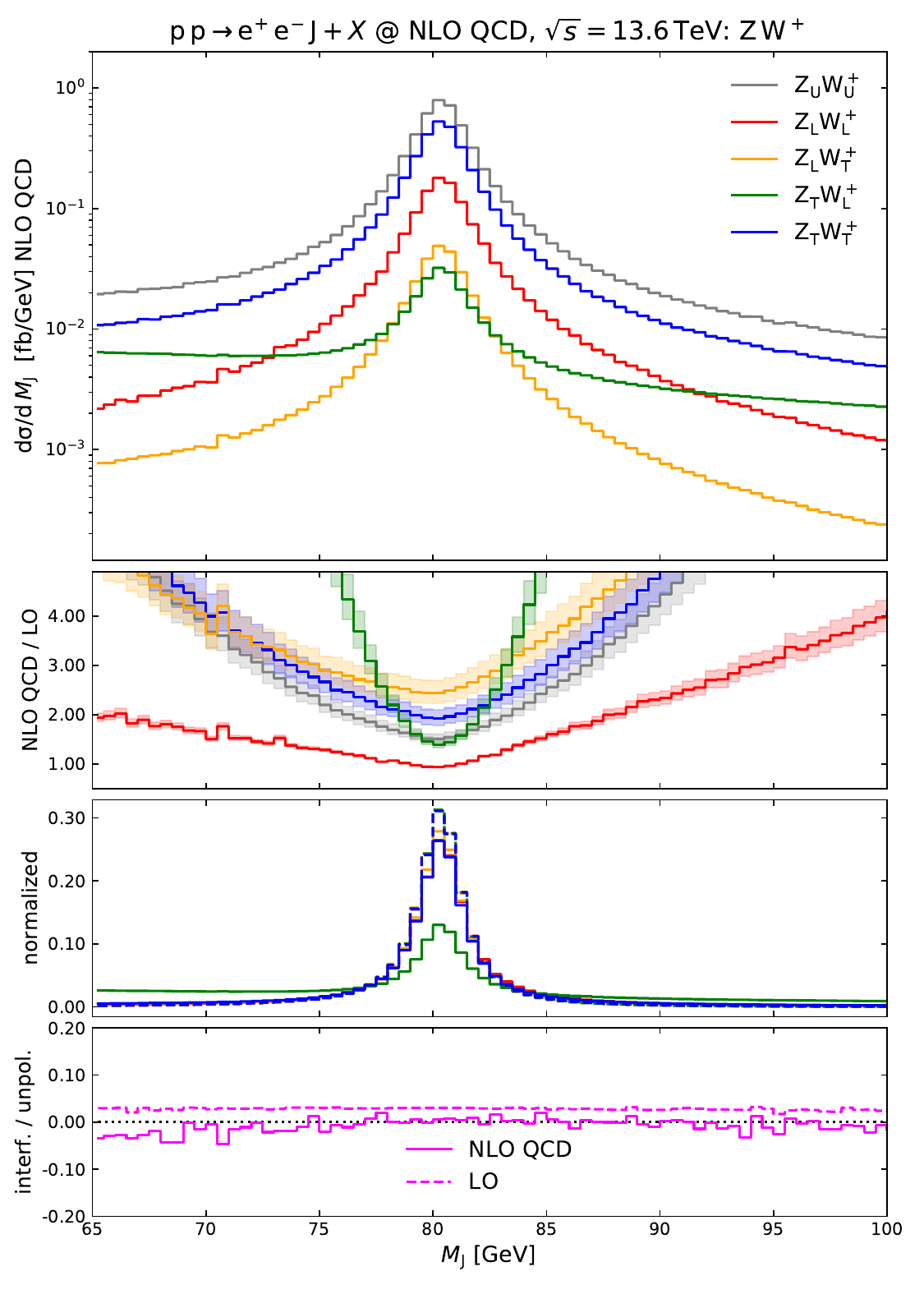}}
  \caption{
    Invariant-mass distribution of the hadronic system $\rm J$ 
    in semi-leptonic $\PZ\PW^+$~production at the LHC.
    The identification of the hadronic system  $\rm J$ is described in \refse{subsec:setupscale}.
    Same structure as \reffi{fig:dec}.
  }\label{fig:mj} 
\end{figure}

Although the angular observables are the most promising ones
in terms of the discrimination power amongst polarisation states,
it is important to complement their investigation with the study of
energy-dependent observables.
In \reffi{fig:mj} we show differential results in the invariant mass of the hadronic system~$\rm J$.
While the distributions feature, as expected, the Breit--Wigner shape
of the $\PW$-boson resonance, this observable is heavily
affected by the reconstruction of the hadronic decay and therefore
subject to contamination from QCD radiation. In fact, the
NLO QCD real corrections introduce an ambiguity in the determination
of the $\PW^+$-boson decay jets.  The jets that happen to be part
of the reconstructed hadronic system  $\rm J$ without being actual decay
products of the $\PW^+$~boson create a background that does not follow
the Breit--Wigner modulation. This effect is particularly manifest 
in the TL distribution. When the recombined $\PW$~boson becomes off-shell, the TL
curve does not fall off like the others (and in particular the LT one,
which is almost identical at LO), but rather results in a much flatter
behaviour.
The striking difference between the TL and LT states, beyond the
different overall normalisation that has already been motivated in
\refse{subsec:int_res}, is due to the interplay between the
reconstruction of the hadronic system and the suppression of
longitudinal bosons in mixed polarisation states in the high-energy
regime.  In particular, both states are unitarity suppressed at LO (as
can be easily seen upon replacement of the longitudinal boson with the
corresponding would-be Goldstone boson), while at NLO QCD there is an
enhancement due to the opening of the gluon-(anti)quark channel that
gives a sizeable real correction. As discussed in
\refse{subsec:int_res}, for the LT polarisation state the
bremsstrahlung parton is produced preferably opposite to the
direction of the hadronically decaying $\PW$~boson.  Therefore little
ambiguity is left for the assignment of decay jets to the transverse
$\PW$~boson.  In the TL case, on the other hand, the additional parton
is produced close to the longitudinal, hadronically decaying $\PW$
boson. This deteriorates the reconstruction of the $\PW$ decay
products and results in a distorted shape with a Breit--Wigner peak
(events with correct assignment, lower in shape compared to the LT
state) on top of a sizeable flat background (events with wrong
assignment). This feature originates from the hard $p_{\rT}$ cut that is
applied to the hadronically decaying $\PW$~boson, vetoing effectively
jets with soft and moderate transverse momentum.

\begin{figure}
  \centering
  \subfigure[$M_{\rm J}$             \label{fig:loose_mj}]{\includegraphics[scale=0.39]{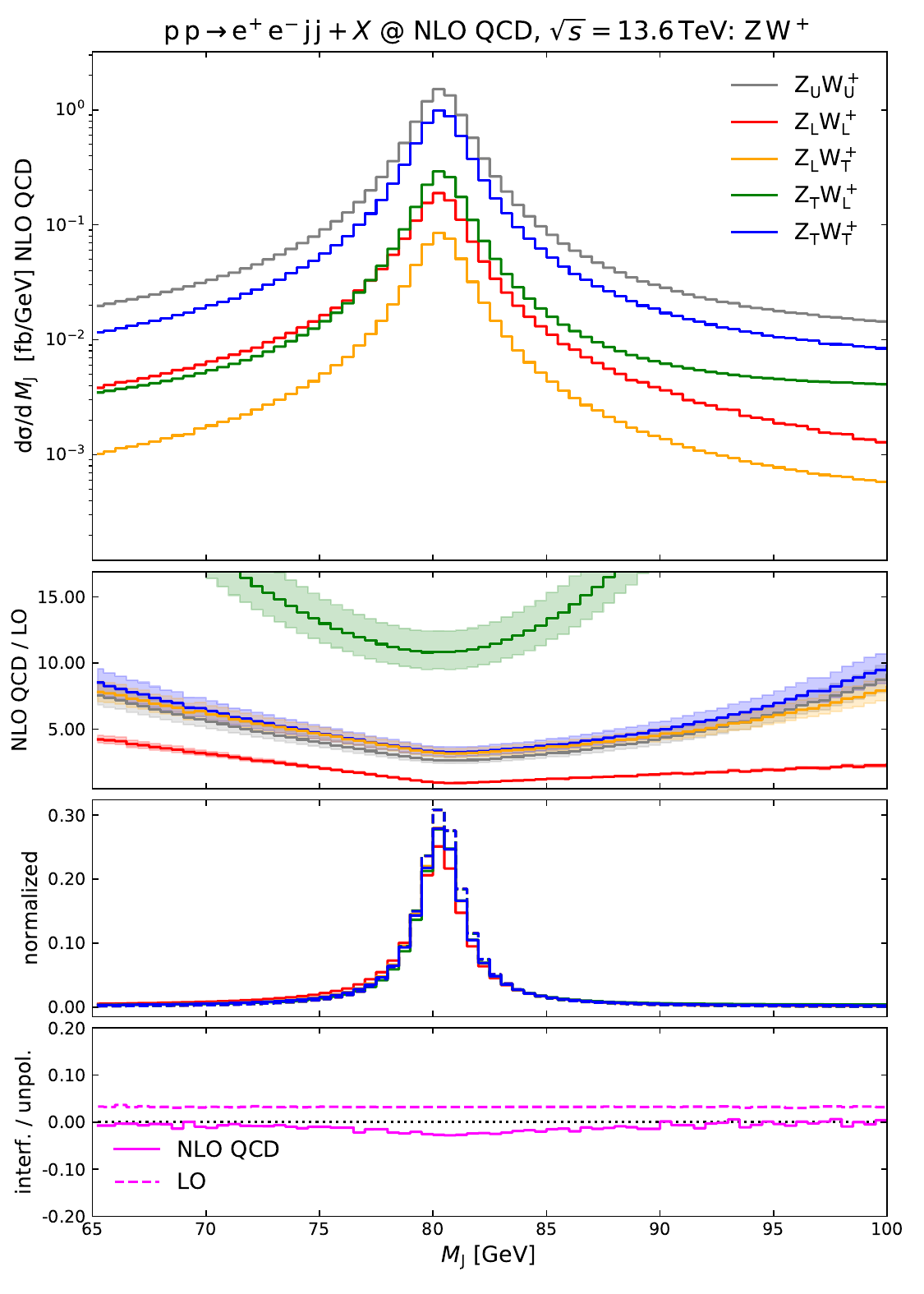}}
  \subfigure[$\pt{\Pj_1}$ \label{fig:loose_ptj1}]{\includegraphics[scale=0.39]{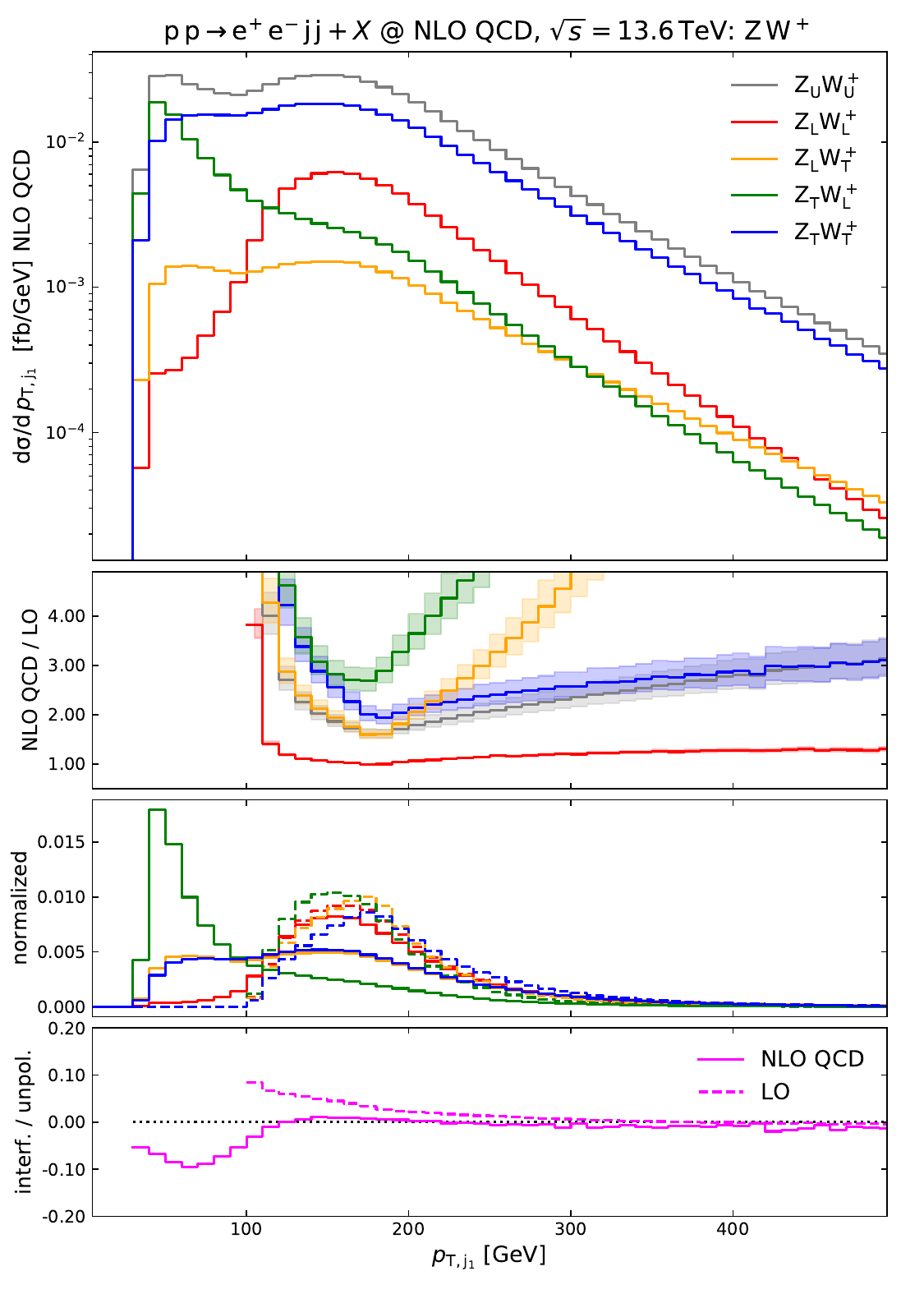}}
  \caption{
    Distribution in the invariant mass of the hadronic system $\rm J$ (a)
    and in the transverse momentum of the hardest decay jet (b)
    in semi-leptonic $\PZ\PW^+$~production at the LHC.
    The identification of the hadronic system  $\rm J$ is described in
    \refse{subsec:setupscale}. The resolved setup is considered,
    but no minimum cut is applied on $\pt{\Pj\Pj}$.    
    The panels of the subfigures have the same structure as in \reffi{fig:dec}.    
  }\label{fig:mj_loose} 
\end{figure}

We have checked that without the hard $p_{\rT}$ requirement on the
hadronic system~$\rm J$  the shape of the TL distribution is closer to the
others, as can be appreciated in \reffi{fig:loose_mj} where we
consider the resolved setup as in \reffi{fig:mj_res} but without the
$\pt{\Pj\Pj}>200\GeV$ cut.
The assignment of decay jets to the $\PW$~boson, although not perfect,
behaves much better for the TL mode, further confirming the reasoning above and in
particular the correlation between the flat background
from mis-reconstruction and the high-$\pt{\rm J}$ cut.
The absence of a hard transverse-momentum cut on the hadronic decay system~$\rm J$
leads to abundant real QCD radiation filling the pure radiative region
below $100\GeV$ for the leading decay jet, which is 
excluded at LO by the required $p_{\rT}$ of the $\PZ$~boson, as can be observed in
the distribution in the $p_{\rT}$ of the hardest
jet from the $\PW$-boson decay shown in \reffi{fig:loose_ptj1}. The distribution
for the TL mode exhibits a clear peak around $50\GeV$, while
a hard cut $\pt{\Pj\Pj}>200\GeV$ would remove all events with
$\pt{\Pj_1}<100\GeV$. The same effect is present also for the LT and
TT polarised states, but much less pronounced. The interference
contribution is a the level of $10\%$ for $\pt{\rm J}<100\GeV$.

Coming back to the results in the default setups shown in \reffi{fig:mj}, the differences
in the TL distribution between the resolved and unresolved topology
are due to the effects of the clustering algorithms on the
reconstruction procedure.  In particular, the flat background from
mis-reconstruction of the $\PW$-boson decay jets is larger (giving a normalised shape
with a lower peak) in the unresolved setup, because the larger
recombination radius causes more initial-state QCD radiation fall in the decay-jet
system.
Looking at \reffi{fig:mj}, the TT and LT distributions feature very
similar shapes when going far from the on-shell regime, since the QCD
effects do not depend much on the polarisation mode of the leptonic
$\PZ$~boson, apart from the different normalisation determined by the
unitarity suppression of the longitudinal polarisation.  The LL state
is characterised by QCD $K$-factors that are below one at the peak,
while they are monotonically increasing in off-shell regimes. In the
resolved setup, the corrections increase faster for $M_{\rm J}<\MW$,
resulting from events where one of the decay jets is missed in the
reconstruction of the $\PW$~boson, while in the unresolved one the
corrections are larger for $M_{\rm J}>\MW$, indicating an
initial-state-radiation jet
be clustered together with one of the decay jets. This is a direct
consequence of the more inclusive jet clustering in the unresolved
setup.  For the TT, LT, and LL states the omission of the
$\pt{\Pj\Pj}$ cut does not change the general picture, apart from the
different normalisation.

\begin{figure}[tb]
  \centering
  \subfigure[Resolved topology \label{fig:mej_res}]  {\includegraphics[scale=0.39]{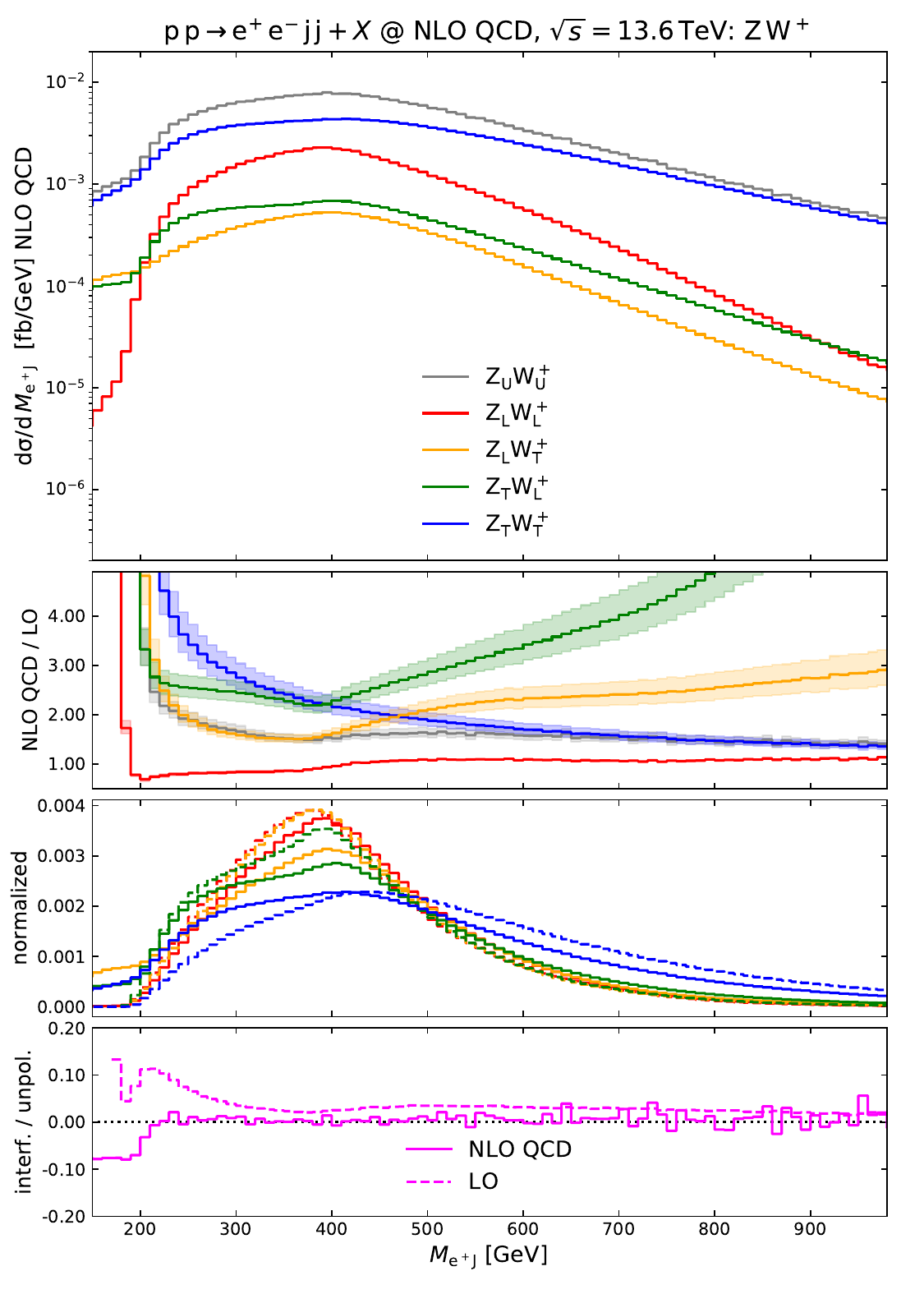}}
  \subfigure[Unresolved topology \label{fig:mej_unres}]{\includegraphics[scale=0.39]{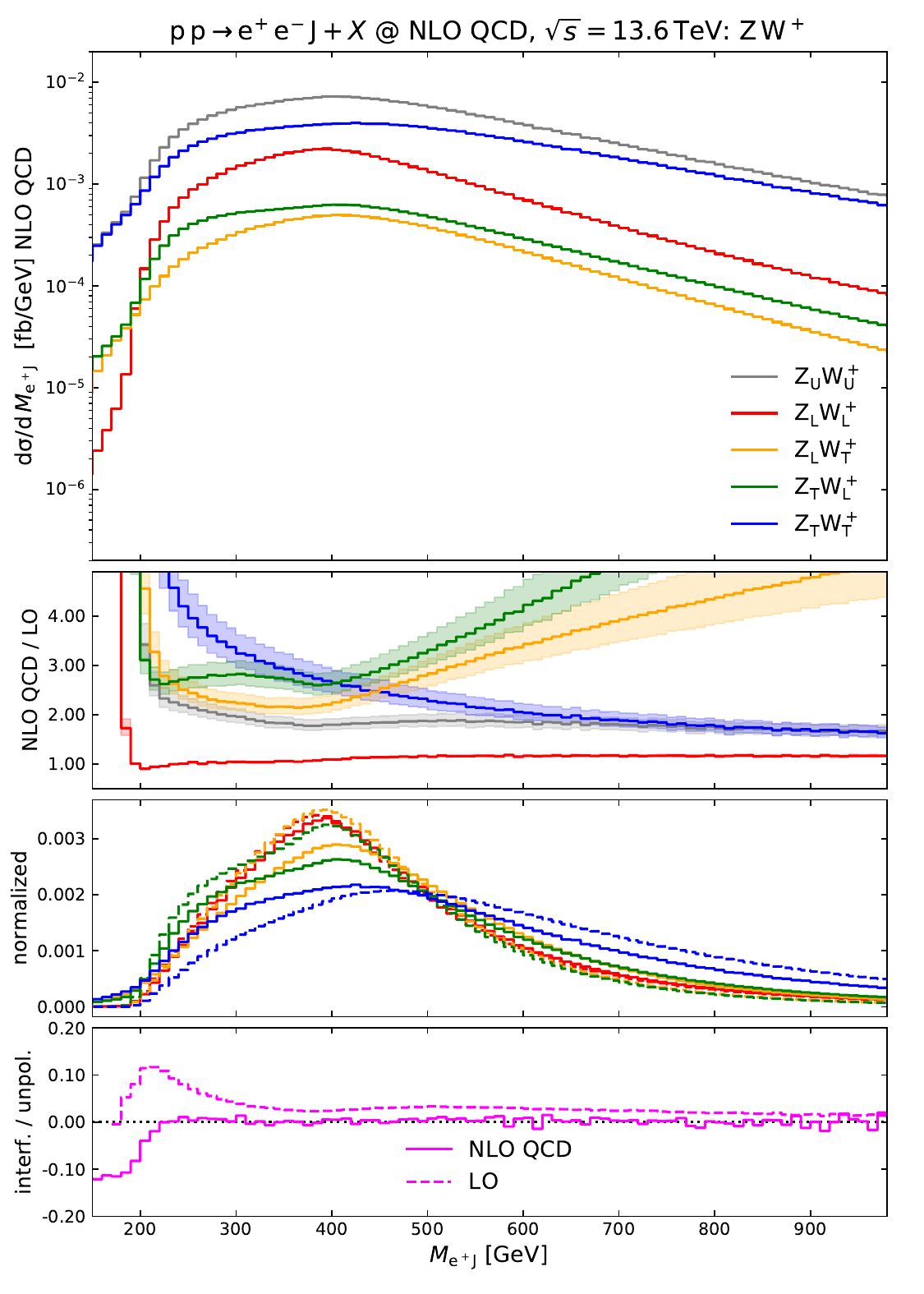}}
  \caption{
    Invariant-mass distributions of the system formed by the
    hadronic system $\rm J$ and the positron 
    in semi-leptonic $\PZ\PW^+$~production at the LHC.
    The identification of the hadronic system  $\rm J$ is described in \refse{subsec:setupscale}.
    Same structure as \reffi{fig:dec}.
  }\label{fig:mej} 
\end{figure}

In \reffi{fig:mej} the differential cross-section is presented with
respect to the invariant mass of the hadronic system  $\rm J$ and the positron.
Large differences amongst various polarisation states occur.
The low-invariant-mass regime ($M_{\Pe^+ \rm J} < 200 \GeV$)
is dominated by real-emission contributions.  In fact,
the additional QCD jet allows for the $\PZ$ and $\PW^{+}$ bosons to be
produced with a lower boson-pair invariant mass. The larger
jet-recombination radius in the unresolved setup leads to higher
$M_{\Pe^+ \rm J}$ thus suppressing the contributions in the
low-invariant-mass region.
At large $ M_{\Pe^+ \rm J}$ the TT distribution falls off slower than all the others. 
In the resolved setup the LL distribution becomes lower than the TL one around $900\GeV$,
while in the unresolved setup the LL signal remains larger than TL, with similar suppressions
at high energy. The much stronger suppression of the LL state in the resolved setup
leads to a signal that is almost one order of magnitude smaller than in the unresolved case
at $1\TeV$. This is due to the fact that for the LL state the high partonic energy is
shared between the two bosons (effect of additional radiation is small for the LL state) resulting
in LO-like configurations with two collimated quarks that are easily clustered
together. This clearly disfavours the resolved topologies where two jets are required.
Rather large negative interferences are present in the radiation-driven soft
part of the spectrum at NLO ($\approx -10\%$).
Shape-wise the LL distribution features a pronounced peak around $400\GeV$,
while the other polarisation states, and especially the TT one, are more spread
over the shown spectrum.\\

\subsubsection{Transverse-momentum distributions}
\begin{figure}[tb]
  \centering
  \subfigure[Resolved \label{fig:ptep_res}]{\includegraphics[scale=0.39]{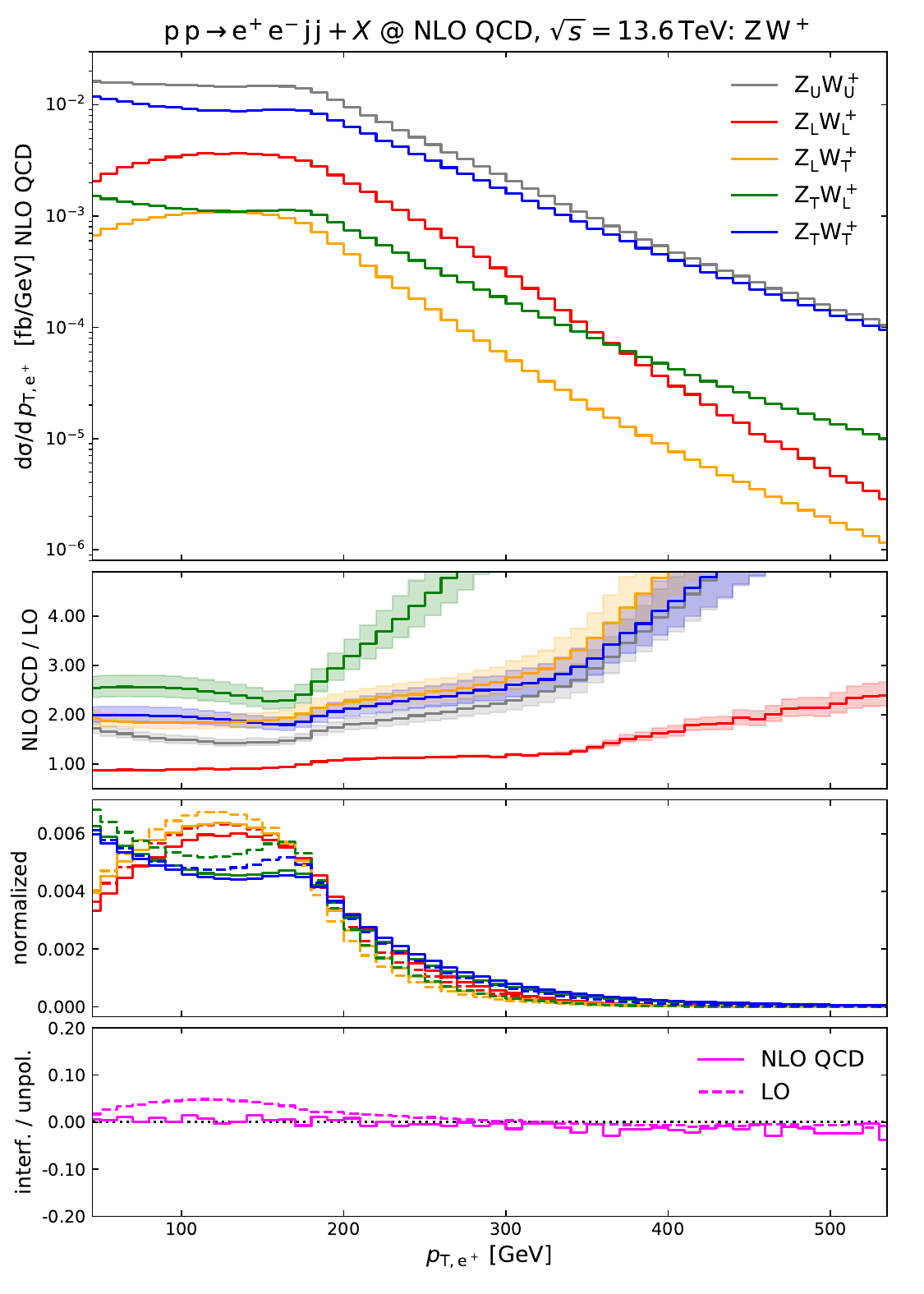}}
  \subfigure[Unresolved topology \label{fig:ptep_unres}]{\includegraphics[scale=0.39]{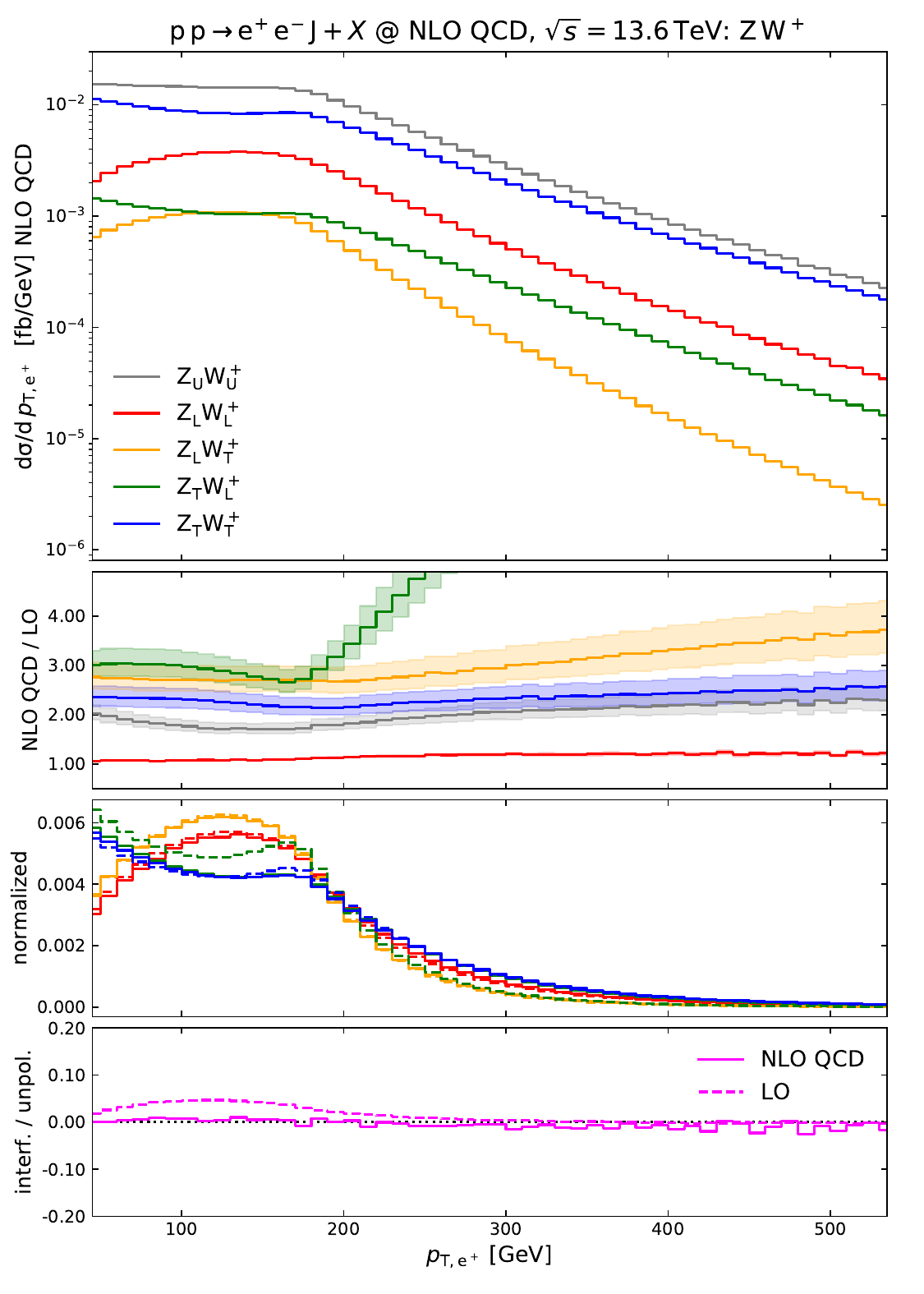}}
  \caption{
    Transverse-momentum distributions of the positron
    in semi-leptonic $\PZ\PW^+$~pro\-duc\-tion at the LHC.
    Same structure as \reffi{fig:dec}.
  }\label{fig:ptep} 
\end{figure}

Similar features are found in transverse-momentum distributions for the positron, which are
shown in \reffi{fig:ptep}.
The behaviour in the low transverse-momentum region strongly depends
on the polarisation of the $\PZ$~boson (of which the positron is a
decay product) and affects the normalised shapes. For a transverse
(longitudinal) $\PZ$~boson the shape has a local minimum (maximum)
around $\pt{\Pe^+}\approx 130 \GeV$.  This is due explained looking at
\reffi{fig:dec_res}: a transverse polarisation gives a
positively-charged lepton that goes more frequently in the same or
opposite direction w.r.t.\ the $\PZ$~boson, therefore sharing in a non-democratic
way the boson energy (two peaks at $\pt{\Pe^+}\approx 20
\GeV$ and $\pt{\Pe^+}\approx 200 \GeV$), while a longitudinal one
gives leptons that are preferably orthogonal to the boson trajectory and share
democratically the boson energy (single peak at $\pt{\Pe^+}\approx 130
\GeV$).
The marked shape differences and the small interference effects
for $\pt{\Pe^+}\lesssim 200\GeV$, namely in the most-populated region,
makes this transverse-momentum observable suitable for the discrimination
of the $\PZ$-boson polarisation state.
Another interesting aspect of the results is the slower fall-off of
the distributions in the unresolved setup compared to the resolved
setup, which is particularly significant for the double-longitudinal
signal. This is the same effect as observed in \reffi{fig:mej} for the
invariant mass of the positron--jet system, which is highly correlated
to the transverse momentum of the positron.  For $\pt{\Pe^+}\gtrsim
300\GeV$, all $K$-factors increase faster in the resolved topology
compared to the unresolved one, owing to  a different LO
suppression. Requiring at least two jets in the final state results in
high-$p_{\rT}$ events being cut away at LO, as the almost collinear
decay quarks are often clustered into a single jet.  In the unresolved
topology such events are not discarded (at least one fat jet is required).
Since the two vector bosons are produced with opposite transverse
momenta, the same effects are found in the high-energy tails of the
transverse-momentum distributions for the hadronic system~$\rm J$, the
$\PZ$~boson, and the charged leptons.\\

\subsubsection{Distributions depending on individual decay jets in the resolved topology}

\begin{figure}[tb]
  \centering
  \subfigure[ \label{fig:thetastar_jet1_res}]{\includegraphics[scale=0.39]{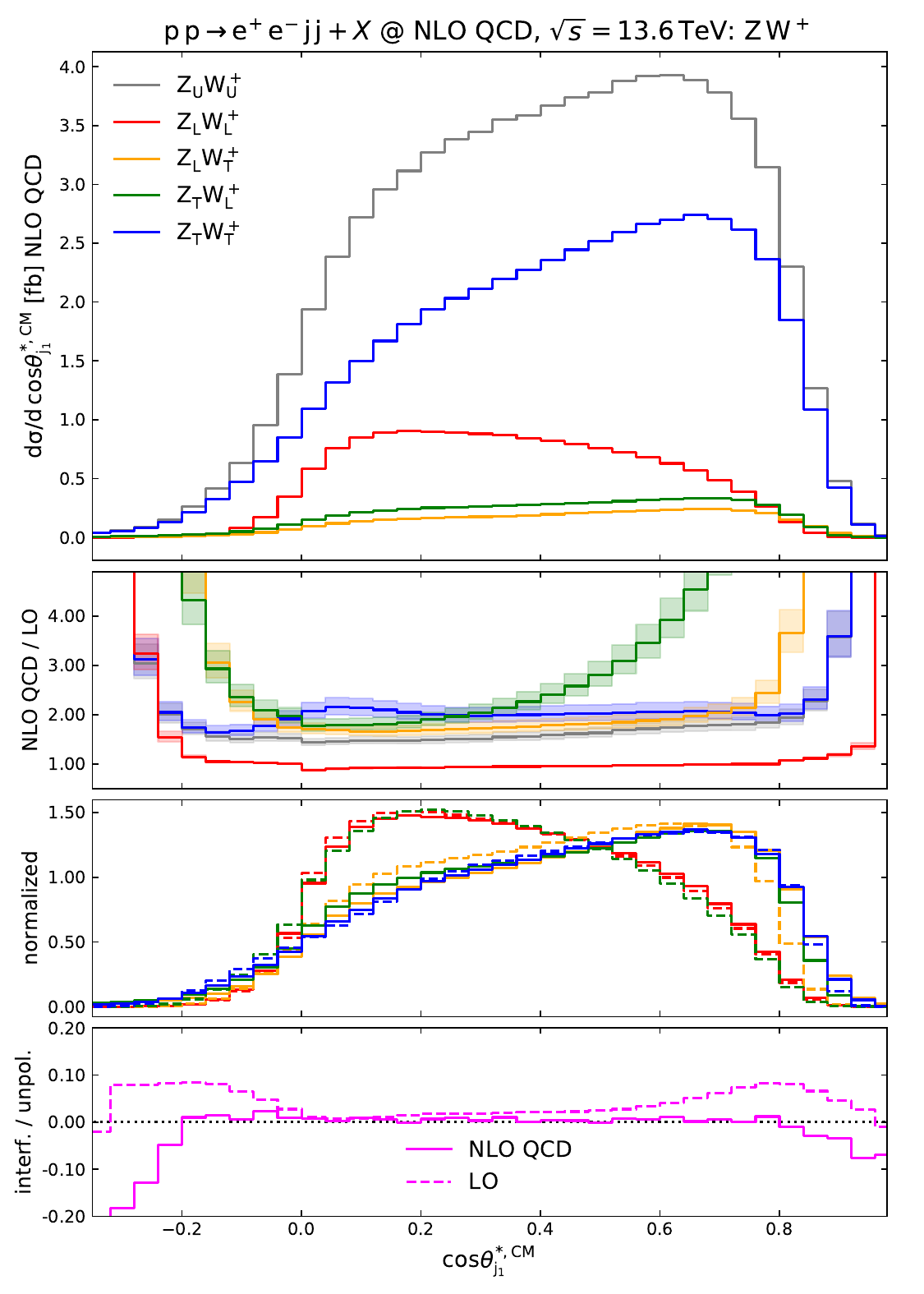}}
  \subfigure[ \label{fig:pt_jet2_res}]{\includegraphics[scale=0.39]{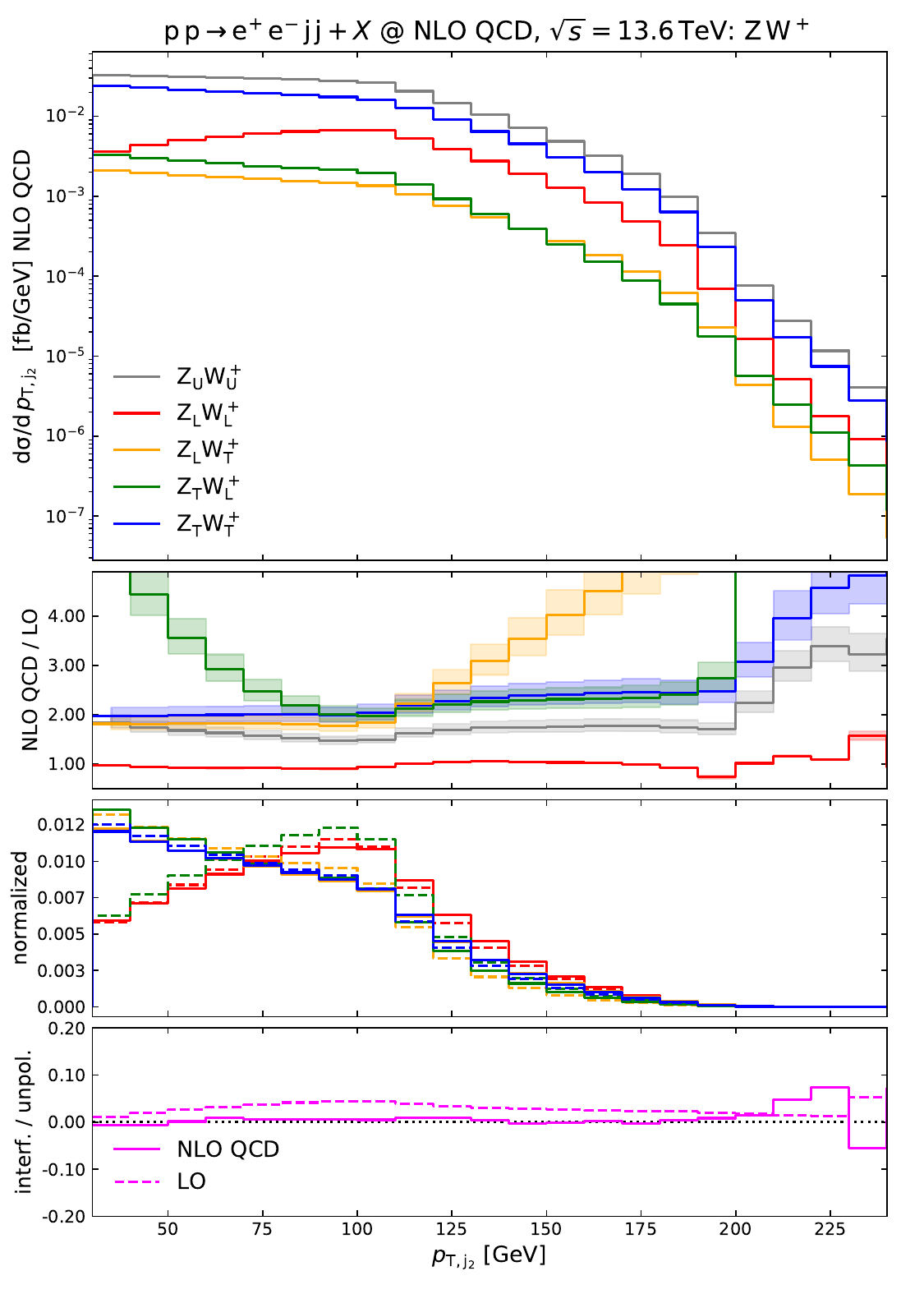}}
  \caption{
    Distributions in the leading-jet decay angle (left)
    and in the subleading-jet transverse momentum (right)
    in semi-leptonic $\PZ\PW^+$~production at the LHC.
    The identification of the leading and subleading jet
    is discussed in \refse{subsec:setupscale} and the decay-angle
    definition is given in Eq.~\eqref{eq:thetastarj}.
    Results for the unpolarised and doubly-polarised
    process are shown in the resolved setup described
    in \refse{sec:kinsetups}.
    The panels of the subfigures have the same structure as in \reffi{fig:dec}.    
  }\label{fig:resonly} 
\end{figure}

In the resolved topology it is possible to distinguish the two jets
that come from the $\PW$-boson decay, up to potentially relevant
reconstruction effects. In \reffi{fig:resonly} we consider two
observables that depend on the kinematics of individual jets (those
labelled as decay jets, sorted according to their transverse
momentum).
The polar decay angle of the leading decay jet $\Pj_1$ in \reffi{fig:thetastar_jet1_res}
is defined similarly to the one of the charged lepton in Eq.~\eqref{eq:thetastar}: it is the
angular separation between the leading-jet direction in the rest frame
of the hadronic system~$\rm J$ ($\vec{p}^{*}_{\Pj_1}$) and the direction of $\rm J$ calculated in the reconstructed di-boson CM frame ($\vec{p}^{\,\CM}_{\rm J}$),
\beq\label{eq:thetastarj}
\cos\theta^{*,\rm CM}_{\Pj_1}\,=\,\frac{\vec{p}^{*}_{\Pj_1} \cdot \vec{p}^{\,\CM}_{\rm J}}{|\vec{p}^{*}_{\Pj_1}||\vec{p}^{\,\CM}_{\rm J}|}\,.
\eeq
Note that this observable is sensitive to the reconstruction procedure to
identify the hadronic system~$\rm J$ and is not selecting the up-like or
down-like jet (which would be unphysical) but rather the leading-$p_{\rT}$ one.
At variance with the leptonic decay angle in the $\PZ$-boson
rest frame, the hardest jet can originate from an up-type or a down-type quark,
strongly distorting the description of the boson decay.
Strikingly, all distributions are non vanishing only between
$\cos\theta^{*,\rm CM}_{\Pj_1}\approx -0.4$ and $\cos\theta^{*,\rm
  CM}_{\Pj_1}=1$, clearly favouring the positive region of the
spectrum.  In fact, the hardest jet is mostly produced in the same
direction of the decayed boson, of which it takes the largest fraction
of transverse momentum. The analogous distributions for the softest
jet show the opposite behaviour, populating mostly the negative region
of the spectrum.  The LO shape for the TL state follows the one of the
LL state. 
This is expected as this angular variable is directly related
to the polarisation state of the $\PW$~boson (and agnostic to the
$\PZ$-boson one). This relation is
somewhat deteriorated, compared to the decay into leptons, since the
identification of the flavour and charge of individual quarks is not
physical, therefore the decay jets can only be ordered according to
their transverse momenta.
In contrast, at NLO QCD the TL shape is following the LT and TT ones.
This dramatic change in the TL shape, driven by large and non-flat
real QCD corrections in gluon-induced channels, is due to the
combination of the bad reconstruction of the $\PW$~boson, discussed
already for \reffis{fig:mj} and \ref{fig:mj_loose}, the suppression of
the LO signal, and the choice of sorting the decay jets in $p_{\rT}$.
The interference effects at NLO QCD are practically negligible in the
most populated region, while they increase up to more than $10\%$
towards the endpoints of the spectrum.

In \reffi{fig:pt_jet2_res} we show the differential cross-section with
respect to the transverse momentum of the softest jet from the
$\PW$-boson decay. As expected, all distributions decrease very fast
already at moderate transverse momentum.  For the LL polarisation state
(both at LO and at NLO) and the TL one (just at LO), the distributions
are peaked around
$100\GeV$, namely half of the minimum transverse momentum required by
the selections for the $\PW$~boson. This behaviour is understood as
the longitudinal $\PW$~boson, mostly produced at large scattering angles
(see \reffi{fig:scatt}),
favours configurations where the two decay jets are orthogonal to the
boson direction (in its rest frame) and therefore typically share
half of the boson transverse momentum.
At NLO, the LL shape does not deviate from the LO one as the QCD corrections are very
small, while the TL shape becomes peaked around zero, following closely the
shape of the TT and LT distributions. This strong modification of the TL
distribution is due to large effects of mis-reconstruction induced by real QCD
radiation.
In particular, the gluon-induced contributions that dominate the NLO QCD corrections
to the TL state, shown in \reffi{subfig:ug2}, prefer a boosted, transversely
polarised $\PZ$~boson whose recoil is absorbed by the system of a hard QCD parton
and a soft pair of quarks from the longitudinal-$\PW$-boson
decay. After clustering and reconstruction, this topology
results in one hard jet and one
soft jet, mimicking the jet pattern characteristic for the hadronic
decay of a transverse $\PW$~boson (TT and LT modes).
The TT and LT distributions show
very similar shapes, dominated by the soft region
($\pt{\Pj_1}<100\GeV$), while for $\pt{\Pj_1}>100\GeV$ the QCD
corrections enhance the LT signal. In fact, for transverse $\PW$~bosons, one
decay jet is preferably emitted opposite to the direction of the
boson, resulting in a small transverse momentum of this subleading
decay jet.

We summarise our main findings for distributions. On the one hand,
the NLO QCD corrections to the distribution in the
leptonic-decay angle of the $\PZ$~boson follow closely those for the
fiducial cross-section with mild shape modifications. On the other hand,
QCD corrections sizeably
distort the shapes of the distributions in the scattering angle and
in the rapidity difference between the positron and the hadronic
system. The radiative corrections significantly change the shape of the
rapidity distribution of the hadronic system for final states with
two longitudinal vector bosons. For the TL polarisation state, the
NLO QCD corrections deteriorate the reconstruction of the
hadronically decaying $\PW$~boson resulting in a flat non-resonant
background in the corresponding invariant-mass distribution,
especially in the presence of a strong cut on the transverse momentum of the
reconstructed $\PW$~boson. The QCD corrections and their interplay with the
employed reconstruction technique lead to a distortion of the distribution
in the invariant mass of the positron and the hadronic system, as well as
the distributions involving resolved decay jets, in particular for the TL
polarisation state.

Besides decay-angle distributions, also the distributions in the
scattering angle and in the rapidity difference between the positron
and the hadronic system are sensitive to the polarisations of the
vector bosons, although the latter two are model dependent. Further
distributions that are sensitive to different vector-boson
polarisations are those in the transverse momentum of the positron
and in the invariant mass of the positron and the hadronic system.
Distributions depending on the decay jets are sensitive to the
polarisation of the hadronically decaying W boson, although they can
be heavily distorted by NLO QCD corrections. Whether the sensitivity
to the polarisation is still present in distributions after a parton
shower matching is an open question.\\

\section{Conclusion}\label{sec:con}
In this work we have presented NLO QCD corrections to vector-boson-pair
inclusive production in the semi-leptonic decay channel. The results
focus on the $\PW\PZ$~process in final states with two charged leptons
and jets, but can be easily extended to $\PZ\PZ$~production with the
same final state, as well as to processes with a charged lepton,
missing transverse momentum and jets ($\PW\PZ$ and $\PW^+\PW^-$). The
building blocks at NLO QCD are exactly the same as those used for this
calculation.

Although we have neglected a number of effects, including the overlap with other production
mechanisms, the non-resonant background and other sources of
corrections (NLO EW, matching to parton shower), the presented calculation
represents a crucial step towards precise predictions for di-boson processes
in semi-leptonic decay channels. For the first time, we have combined
in the double-pole approximation the QCD corrections to the production of two bosons and
to the hadronic decay of one of the two, separating doubly-polarised signals at the level
of tree-level and one-loop Standard Model amplitudes.

We have considered a boosted regime, where the longitudinal signals
give a more sizeable contribution than in inclusive setups. We have
applied two different jet selections: a first one with two light jets
(resolved) and a second one with a single fat jet (unresolved).
Between the two setups moderate differences show up  at the level of
distribution shapes and more marked deviations are revealed for QCD
$K$-factors for the various polarisation states.

The reconstruction of the hadronic decay of the $\PW$~boson is found
to behave very differently for the various polarisation states,
distorting angular and energy-dependent distributions and enhancing
otherwise suppressed contributions. The largest impact 
is found when the $\PW$~boson is longitudinally polarised and
the leptonically decaying $\PZ$~boson is transversely polarised, as a combination
of unitarity cancellations and a hard cut on the transverse momentum
of the longitudinal bosons.
For this polarisation mode, the sizeable QCD corrections and
their interplay with the reconstruction procedure 
causes distributions for decay observables of the longitudinal
$\PW$~boson to mimic those of transverse bosons.

Strikingly, a number of observables turn out to be highly sensitive to
polarisation-state discrimination. Many of these observables are
inclusive in the hadronic decay structure, \ie they do not rely on
jet-substructure techniques.  
This clearly suggests that
  extracting relevant polarisation information from the data is
  possible even avoiding any reconstruction of the sub-jets from the
  hadronic decay.  Resolving individual decay jets from the hadronic
  decay gives access to a larger set of distributions which are
  sensitive to the polarisation of the decaying boson.  These results
  demonstrate that semileptonic final states could give an important
  boost to the investigation of polarised di-boson production, both
  via decay-specific analyes and as complementary to fully-leptonic
  final states.  In spite of very large backgrounds to be subtracted,
  the semi-leptonic channel would in fact enhance the sensitivity to
  weak-boson polarisations, otherwise restricted to fully-leptonic
  final states, which are clean but statistically limited.

The dramatic change of (doubly-)polarised distributions when going from LO to NLO QCD
makes it essential to include at least NLO QCD corrections in any polarisation study
or data analysis in the considered decay channel. This statement, however,
applies also for fully-leptonic channels, as shown in previous works
\cite{Denner:2020bcz,Denner:2020eck,Denner:2021csi,Le:2022lrp,Le:2022ppa,Poncelet:2021jmj}.
Including NLO corrections is especially important for multi-boson processes
that are characterised by a LO suppression in some kinematic configurations.
The inclusion of NNLO QCD corrections, though definitely desirable and now
feasible for di-boson processes \cite{Poncelet:2021jmj}, is not expected to give
as dramatic shape distortions to the NLO distributions, as those
given by the NLO corrections to the LO shapes.
With specific regards to the hadronic decays of EW bosons, it will be especially
relevant to match NLO QCD (or NNLO QCD) calculations to parton showers and hadronisation,
enabling a realistic comparison against LHC data. 
It is hard to estimate the impact that the parton-shower
matching will have on the sensitivity to vector-boson polarisations, although
in general a deterioration could be expected owing to additional radiations
coming both from the initial state and from the boson decay.
A public code to match  polarised-boson fixed-order calculations and
parton showers
is still lacking. However, since the external degrees of freedom of the
matrix elements are unpolarised, even in the presence of intermediate vector bosons
with fixed polarisation state, we do not foresee conceptual complications in
the matching of polarised fixed-order results to parton-shower effects
  compared to the case of unpolarised vector bosons.

\section*{Acknowledgements}
The authors are grateful to Jean-Nicolas Lang for maintaining \recola and to Stefan Rode for testing one-loop 
amplitudes with intermediate polarised bosons.
We would like to thank Lucia Di Ciaccio, Joany Manjarres,  Mathieu Pellen, Ren\'e Poncelet, Emmanuel Sauvan and Frank Siegert
for useful discussions.
This work is supported by the German Federal Ministry for Education and Research
(BMBF) under contract no.~05H21WWCAA and by the German Research Foundation
(DFG) under reference number DFG 623/8-1.

\bibliographystyle{JHEPmod}
\bibliography{polvv}

\end{document}